\documentclass[english]{aa}
\usepackage{times}
\usepackage[T1]{fontenc}
\usepackage[latin1]{inputenc}
\usepackage{amsmath}
\usepackage[dvips]{graphicx}
\usepackage{amssymb}
\usepackage[authoryear]{natbib}

\makeatletter

\newcommand{\noun}[1]{\textsc{#1}}
\providecommand{\tabularnewline}{\\}

\usepackage{txfonts}
\newcommand{\Teff}{T_{\rm eff}}
\newcommand{\Menv}{M_{\rm env}}
\newcommand{\Msol}{M_\odot}

\authorrunning{Charpinet et al.}
\titlerunning{Structural Parameters of the Hot Pulsating B Subdwarf
\object{PG~1219+534} from Asteroseismology}

\usepackage{babel}
\makeatother

\begin{document}

\title{Structural Parameters of the Hot Pulsating B Subdwarf
\object{PG~1219+534} from Asteroseismology\thanks{Based on photometric observations gathered at the Canada-France-Hawaii Telescope, operated by the National Research Council of Canada, the Conseil National de la Recherche Scientifique of France, and the University of Hawaii. Spectroscopic observations reported here were obtained at the MMT Observatory, a joint facility of the University of Arizona and the Smithsonian Institution.}$^{,}$\thanks{This study made extensive use of the computing facilities offered by the Calcul en Midi-Pyr\'en\'ees ({\sc Calmip}) project, France.}}

\author{S. Charpinet\inst{1}\and G. Fontaine\inst{2}\and P. Brassard\inst{2}\and
E.M. Green\inst{3}\and P. Chayer\inst{4, 5}}

\institute{UMR 5572, Universit\'e Paul Sabatier et CNRS, Observatoire Midi-Pyr\'en\'ees,
14 Av. E. Belin, F-31400 Toulouse, France\\
\email{scharpin@ast.obs-mip.fr}\and D\'epartement de Physique, Universit\'e de Montr\'eal, C.P. 6128, Succursale Centre-Ville, Montr\'eal QC, H3C
3J7, Canada\\
\email{fontaine,brassard@astro.umontreal.ca}\and Steward Observatory,
University of Arizona, 933 North Cherry Avenue, Tucson, AZ 85721,
USA\\
\email{bgreen@as.arizona.edu}\and Department of Physics and Astronomy,
Johns Hopkins University, 3400 North Charles Street, Baltimore, MD
21218-2686, USA\\
\email{chayer@pha.jhu.edu}\and Primary affiliation: Department of
Physics and Astronomy, University of Victoria, P. O. Box 3055, Victoria,
BC V8W 3P6, Canada}

\offprints{S. Charpinet}

\date{Received 17 January 2005; Accepted 10 March 2005}

\abstract{Over the last several years, we have embarked on a long term effort
to exploit the strong potential that hot B subdwarf (sdB) pulsators
have to offer in terms of asteroseismology. This effort is multifaceted
as it involves, on the observational front, the acquisition of high
sensitivity photometric data supplemented by accurate spectroscopic
measurements, and, on the theoretical and modeling fronts, the development
of appropriate numerical tools dedicated to the asteroseismological
interpretation of the seismic observations. In this paper, we report
on the observations and thorough analysis of the rapidly pulsating
sdB star (or EC14026 star) \object{PG~1219+534}. Our model atmosphere
analysis of the time averaged optical spectrum of \object{PG~1219+534}
obtained at the new Multiple Mirror Telescope (MMT) leads to estimates
of $T_{{\rm {\rm eff}}}=$ 33,600 $\pm$ 370 K and $\log g=5.810\pm0.046$
(with $\log N({\rm He})/N({\rm H})=-1.49\pm0.08$), in good agreement
with previous spectroscopic measurements of its atmospheric parameters.
This places \object{PG~1219+534} right in the middle of the EC14026
instability region in the $\log g-T_{{\rm eff}}$ plane. A standard
Fourier analysis of our high signal-to-noise ratio Canada-France-Hawaii
Telescope (CFHT) light curves reveals the presence of nine distinct
harmonic oscillations with periods in the range $122-172$ s, a significant
improvement over the original detection of only four periods by \citet{1999MNRAS.305...28K}.
On this basis, we have carried out a detailed asteroseismic analysis
of \object{PG~1219+534} using the well-known forward method and assuming
that the observed modes have $\ell\le 3$. Our
analysis leads objectively to the identification of the ($k$, $\ell$)
indices of the nine periods observed in the star \object{PG~1219+534},
and to the determination of its structural parameters. The periods
all correspond to low-order acoustic modes with adjacent values of
$k$ and with $\ell=0$, 1, 2, and 3. They define a band of unstable
modes, in close agreement with nonadiabatic pulsation theory. Furthermore,
the average dispersion between the nine observed periods and the periods
of the corresponding nine theoretical modes of the optimal model is
only $\sim0.6$\%, comparable to the results of a similar analysis
carried out by \citet{2001ApJ...563.1013B} on the rapid sdB pulsator
\object{PG 0014+067}. On the basis of our combined spectroscopic
and asteroseismic analysis, the inferred global structural parameters
of \object{PG~1219+534} are $T_{{\rm eff}}=$ 33,600 $\pm$ 370 K,
$\log g=5.8071\pm0.0057$, $\log M_{{\rm env}}/M_{*}=-4.254\pm0.147$,
$M_{*}=0.457\pm0.012$ $M_{\odot}$, $R/R_{\odot}=0.1397\pm0.0028$,
and $L/L_{\odot}=22.01\pm1.85$. Combined with detailed model atmosphere
calculations, we estimate, in addition, that this star has an absolute
visual magnitude $M_{V}=4.62\pm0.06$ and is located at a distance
$d=531\pm23$ pc (using $V=13.24\pm0.20$). Finally, if we interpret
the absence of fine structure (frequency multiplets) as indicative
of a slow rotation rate of that star, we further find that \object{PG~1219+534}
rotates with a period longer than $3.4$ days, and has a maximum rotational
broadening velocity of $V\sin i\lesssim2.1$ km.s$^{-1}$.\\

}

\maketitle

\section{Introduction}

Hot subdwarf B (sdB) stars are commonly interpreted as low-mass ($\sim0.5$
$\Msol$) core helium burning objects that belong to the so-called
Extreme Horizontal Branch (EHB). Having extremely thin and mostly
inert hydrogen rich residual envelopes ($\Menv<0.02$ $\Msol$), these
stars remain hot ($\Teff\sim$ 22,000 -- 42,000 K) and compact ($\log g\sim5.2-6.2$)
throughout their entire EHB lifetime and eventually evolve toward
the white dwarf cooling sequence without experiencing the Asymptotic
Giant Branch (AGB) and Planetary Nebulae (PN) phases of stellar evolution
(see \citealp{1993ApJ...419..596D}). If the fate of sdB stars is
relatively well understood, their origin -- i.e., how they form -- is
still unclear, however. The main difficulty resides in identifying
how sdB stars manage to remove all but a very small fraction of their
envelope at exactly the same time that they reach a sufficient core
mass for helium flash ignition at the tip of the Red Giant Branch
(RGB). Various formation scenarios have been proposed over the years
to solve this problem, involving either single star evolution with
enhanced mass loss at the tip of the RGB \citep{1996ApJ...466..359D},
or binary evolution including various concurrent channels from common
envelope ejection, stable Roche lobe overflow, to the merger of two
helium white dwarfs (see, e.g., \citealt{2002MNRAS.336..449H,2003MNRAS.341..669H}
and references therein). Interestingly, these authors suggest that
the latter binary evolution scenarios would lead to a somewhat broader
dispersion of stellar masses among sdB stars (between 0.30 -- 0.70
$\Msol$) than is currently assumed from canonical EHB models.

Renewed interest in this phase of stellar evolution has come in part
from the recent discoveries of two new classes of multiperiodic nonradial
pulsators among sdB stars which enable, in principle, the asteroseismic
probe of their internal structure. The rapid sdB pulsators known as
the \emph{V361 Hya} stars, but more commonly referred to as the EC14026
stars, were the first to be observationally detected by colleagues
from the South African Astronomical Observatory (SAAO; \citealp{1997MNRAS.285..640K}).
Interestingly, at the same epoch, their existence was also independently
predicted from theoretical considerations based on the identification
of an efficient driving mechanism for the pulsations \citep{1996ApJ...471L.103C}.
The mechanism uncovered is a $\kappa$-effect due to local overabundances
of heavy metals -- especially of iron -- produced by microscopic diffusion
processes which are at work in the envelope of these stars
\citep{1997ApJ...483L.123C}. Various surveys by the SAAO group, by the
Montr\'eal group (see \citealp{2002ApJ...578..515B}), by a
Norwegian-German-Italian team (see, e.g., \citealp{2002A&A...383..239S}),
and by other teams (see the reviews of \citealp{2001AN....322..387C} and
\citealp{2002rnpp.conf..356K})
rapidly led to additional discoveries, bringing the total of known
EC14026 pulsators to 33 at the time of this writing. The EC14026 stars
tend to cluster at high effective temperatures and surface gravities
(near $\Teff\sim$ 33,500 K and $\log g\sim5.8$), although outliers
indicate a relatively large dispersion of these stars in the $\log g-\Teff$
plane. The periods detected in EC14026 stars typically range from
$\sim100$ s to $\sim200$ s, but can be substantially longer for
class members having lower surface gravities (e.g., 290--600 s for
\object{PG 1605+072}; see \citealp{2001AN....322..387C} and references
therein). Their mode amplitudes usually span a relatively wide range,
but typical values are of several millimagnitudes. In most cases,
the short periods are entirely consistent with low-order, low-degree
radial and nonradial acoustic waves (\citealp{1997MNRAS.285..651S};
\citealt{1997ApJ...483L.123C,2001PASP..113..775C}). However, in
low surface gravity
sdB pulsators, the presence in the observed period spectrum of low-order
$g$-modes or \emph{mixed} modes having both $p$- and $g$-mode properties
cannot be excluded and somewhat complicates the identification of
the true nature of the modes being detected \citep{2002ApJS..139..487C}.
Remarkable similarities exist between appropriate EHB pulsation models
(i.e., our so-called {}``second generation'' models which take radiative
levitation of iron into account) and the most basic properties of
EC14026 pulsators (see the review of \citealp{2001PASP..113..775C}).

The long period sdB variables (the PG1716+426 stars, but sometimes also
referred to as the {}``Betsy stars'' or the {}``lpsdBV stars'')
were discovered only a couple of years ago \citep{2003ApJ...583L..31G}.
PG1716+426 stars oscillate on much longer timescales than their EC14026
counterparts, with periods ranging from $\sim45$ minutes to $\sim2$
hours. This implies that relatively high order $g$-modes are involved.
Moreover, unlike the EC14026 pulsators, these stars populate the
low-temperature/low-gravity
domain of the $\log g-\Teff$ plane where sdB stars are found. Remarkably,
however, the same mechanism responsible for the oscillations in the
EC14026 stars is thought to operate in the long period sdB pulsators
as well, but destabilizing this time high-order, $\ell\gtrsim3$ gravity
modes (\citealp{2003ApJ...597..518F}).

For the asteroseismologist, the existence of these two distinct classes
of pulsators among the hot B subdwarfs is of particular interest,
as $p$-modes and $g$-modes probe different regions of their interiors
(see \citealp{2000ApJS..131..223C}). One can therefore hope to gather
complementary information on EHB stars in general from the study
of these two classes of pulsators. While it is too early to assess
the real asteroseismological potential of the newly discovered PG1716+426
stars, the EC14026 pulsators have already proved to be excellent laboratories
for asteroseismic studies. In the exploratory work of
\citet{2001ApJ...563.1013B}, the acquisition of high signal-to-noise
(S/N) ratio data at the Canada-France-Hawaii
3.6 m Telescope (CFHT) combined with efforts to develop a new objective
global optimization method for asteroseismology led to the first successful
attempt to match \emph{all} the detected periods of a short period
sdB pulsator. These authors succeeded in reproducing \emph{simultaneously}
(at the $\sim1\%$ level) 13 periods extracted from the light curve
of the star \object{PG 0014+067}. This led to the first asteroseismological
determination of the fundamental parameters of an sdB star, such as
its effective temperature, its surface gravity (to a greatly improved
accuracy), its total mass, and the mass of its H-rich envelope. The
derivation of that latter quantity, which is of utmost importance in
the context of Extended Horizontal Branch (EHB) stellar evolution to which sdB
stars are bound, is a pure product of asteroseismology and currently
cannot be measured by any other known technique. \citet{2001ApJ...563.1013B}
have emphasized the important fact that their asteroseismological solution is
entirely consistent with both the atmospheric
parameters ($\Teff$ and $\log g$) derived from independent quantitative
spectroscopy and the mode stability predictions obtained from the
nonadiabatic pulsation calculations based on the so-called second
generation models. Such a consistency between three independent aspects
of the modeling of these stars -- namely model atmosphere computations,
period distribution calculations, and nonadiabatic mode stability
predictions -- certainly adds to the credibility of that asteroseismological
result.

Since the work of \citet{2001ApJ...563.1013B}, we have greatly perfected
our technique and successfully extended such studies to several pulsating
hot B subdwarfs, as well as several white dwarf pulsators (see, e.g.,
\citealt{2003whdw.conf...69C}; \citealt{2003whdw.conf..259B}). In
this paper, we report on the case of the hot B subdwarf \object{PG~1219+534}
(aka \object{V* KY UMa}), a star first recognized as an sdB pulsator
by \citet{1999MNRAS.305...28K}. These authors have identified four
periodicities in the discovery light curve of this star. In this study,
we more than double that number by reaching a total of nine independent
periods detected, based on high S/N ratio light curves obtained during
dedicated observations at the CFHT. This significant improvement at
the level of the photometry added to new improved spectroscopy on
that star allows us to perform a full asteroseismic analysis of
\object{PG~1219+534}. This ultimately leads to the determination of its
fundamental parameters, including its total mass and H-rich envelope mass,
i.e., two key parameters for constraining formation and evolution
scenarios of sdB stars in the future.

In the following sections, we fully report on our combined photometric
and spectroscopic observations of \object{PG~1219+534} (Section 2)
and discuss the detailed analysis of the photometric light curves
obtained for that star (Section 3). Then, we present and discuss in
detail the outcome of the thorough asteroseismic analysis we have
carried out on \object{PG~1219+534} based on these new observations
(Section 4). A summary and conclusions uprising from this analysis
are then provided in Section~5.

\section{Observations}

\subsection{Medium Resolution Spectroscopy}

A reference to \object{PG 1219+534} appears for the first time in
the Palomar-Green catalog of UV excess objects \citep{1986ApJS...61..305G}.
Loosely classified at that time as a subdwarf star with no indication
of its subclass, the available Johnson photometry from \citeauthor{1986ApJS...61..305G}
(\citeyear{1986ApJS...61..305G}; $B=12.41$) and Str{\"o}mgren photometry
from \citeauthor{1992AJ....104..203W} (\citeyear{1992AJ....104..203W};
$y=13.235$, $(b-y)=-0.157$, $(u-b)=-0.265$, $m1=0.097$) indicate
however a relatively bright star with colors typical of an isolated
sdB object.

An estimate for the surface parameters of \object{PG 1219+534} based
on the existing absolute multicolor photometry from the literature
and on dedicated spectroscopy from the 4.2 m William Herschel Telescope
(WHT) with the spectrograph ISIS was first proposed in \citet{1999MNRAS.305...28K}.
Fitting the Balmer lines of the spectrum with profiles calculated
from a grid of LTE H/He model atmospheres \citep{1994ApJ...432..351S},
these authors adopted the values $\Teff=$ 32,800 $\pm$ 300 K and
$\log g=5.76\pm0.04$. However, \citet{1999MNRAS.305...28K} could
not provide a direct fit of the helium lines and relied instead on
comparisons with spectra of other sdB stars to estimate a surface
helium abundance of $\log N({\rm He})/N({\rm H})\sim-1.18$, which
is among the highest ratios encountered for sdB stars. This rough
estimate adds additional uncertainty on the $\Teff$ and $\log g$
values thus derived.

\citet{2000A&A...363..198H} provided a detailed analysis of a Keck HIRES
spectrum of \object{PG 1219+534} using line blanketed NLTE and LTE
model atmospheres to derive the atmospheric parameters, metal
abundances, and rotational velocity of that star. These authors used
several indicators based on either hydrogen Balmer lines and neutral He
lines fitting or He ionization equilibrium, but could not reconcile all
measurements. In particular, they pointed out to a ``helium problem'',
i.e., the inability to fit simultaneously well the hydrogen Balmer
lines, the neutral helium lines, and the HeII 4686 line seen in the
spectrum of \object{PG 1219+534}. This problem seems widespread among
the hotter sdB stars. By forcing a better fit to the helium lines, in
particular to the HeII 4686 line, \citet{2000A&A...363..198H} found that
the solution would now suggest higher effective temperatures and higher
surface gravities than those inferred in their global fits (see their
Table 5 and Figure 5). In order to reflect these uncertainties, the
authors finally adopted values of the atmospheric parameters for
\object{PG 1219+534} with relatively large error bars, namely
$\Teff=$34,300$_{{\rm -1,000}}^{{\rm +2,000}}$ K, $\log g=5.95\pm0.1$,
and $\log N({\rm He})/N({\rm H})=-1.50\pm0.10$. The very
sharp metal lines seen in the Keck HIRES spectrum also allowed these
authors to constrain the projected rotational velocity of \object{PG
  1219+534} by setting an interesting  limit of $V\sin i\lesssim10$ km
s$^{-1}$, indicating that this star is either a slow rotator, an object
seen nearly pole-on, or both.

We must point out here that, at the outset, asteroseismology rules out a
surface gravity as high as log $g$ = 5.95 for PG 1219+534 for the simple
reason that there are not enough theoretical modes available in the
range of observed periods. Only models with lower values of log $g$ have
dense enough period spectra, a very robust result that reflects the high
sensitivity of the pulsation periods on that parameter in sdB stars. A
clear demonstration of this phenomenon is provided in Figure 3 of
Fontaine et al. (1998; see also Figure 5.13 of Charpinet 1999).
Accordingly, it will not be surprising to find out below that models
with log $g$ = 5.95 provide extremely poor fits to the period data and
that things get worse when increasing further the surface gravity.
Hence, among the possible atmospheric solutions proposed by
\citet{2000A&A...363..198H}, only those with the lower surface gravities
appear compatible with the period data for PG 1219+534.

\begin{figure}
\begin{center}\includegraphics[%
  clip,
  scale=0.50]{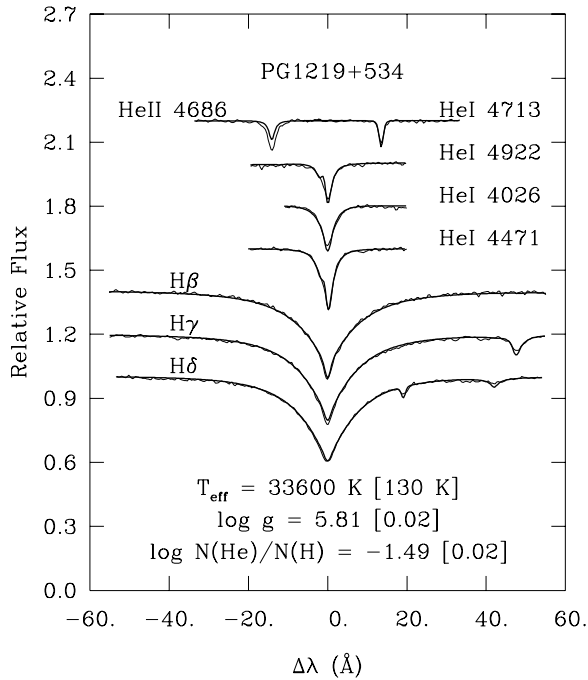}\end{center}

\caption{Model fit (\emph{solid curve}) to the hydrogen Balmer lines
and helium lines available in our time averaged high S/N ratio,
medium-resolution optical spectrum of PG 1219+534\label{cap:spectro1}}
\end{figure}

\begin{figure}
\begin{center}\includegraphics[%
  clip,
  scale=0.44]{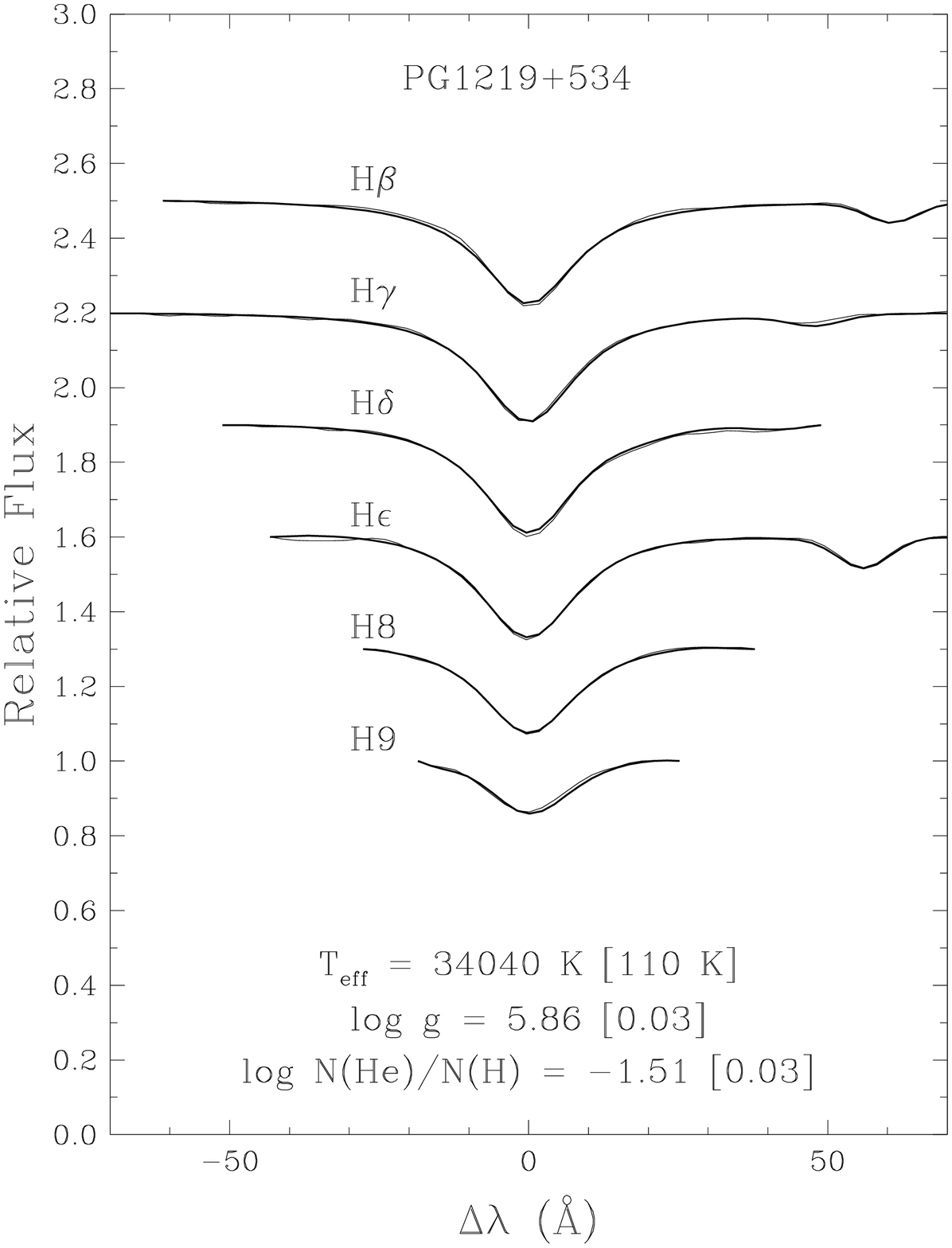}\end{center}

\caption{Model fit (\emph{solid curve}) to the hydrogen Balmer lines
and neutral helium lines available in our time averaged high S/N ratio, low-resolution
optical spectrum of PG~1219+534\label{cap:spectro2}}
\end{figure}

In order to double check on the spectroscopy of the star \object{PG 1219+534},
we obtained additional measurements from which we derived independent
estimates for its atmospheric parameters. A medium-resolution ($\sim1${\AA}),
high S/N ratio ($\sim 175$ per pixel, with 3.1 pixels per resolution
element) spectrum of that star was obtained recently with the blue
spectrograph at the new 6.5 m Multiple Mirror Telescope (MMT). That
spectrum covers the range from $\sim 4000${\AA} to $\sim 4950${\AA}. Another
low-resolution ($\sim 8.7${\AA}), high S/N ratio ($\sim 231$ per pixel, with
2.6 pixels per resolution element) spectrum from Steward's 2.3 m
Telescope was also available. The latter covers the range from
$\sim 3615${\AA} to $\sim 6900$\AA. Such observations are part of a global
program aimed at improving the spectroscopic characterization of sdB
stars, and in particular of EC14026 stars and long period sdB pulsators
(the PG 1716+426 stars).

An integral part of that program is the development of a bank of model
atmospheres and synthetic spectra suitable for the analysis of the
spectroscopic data. To this end, we have so far computed two detailed
grids (one in LTE and the other in NLTE) with the help of the public
codes \noun{Tlusty} and \noun{Synspec} (\citealp{Hubeny1995}, \citealp{Lanz1995}).
Each grid is defined in terms of 11 values of the
effective temperature (from 20,000 K to 40,000 K in steps of 2,000 K),
10 values of the surface gravity (from $\log g$ of 4.6 to 6.4, in steps
of 0.2 dex), and 9 values of the helium-to-hydrogen number ratio (from
$\log N({\rm He})/N({\rm H})$ of $-4.0$ to 0.0, in steps of 0.5 dex). These grids were
originally developped to analyze our MMT data and, therefore, our
current synthetic spectra are limited to the range from 3500{\AA} to
5800{\AA}, but this can easily be widened as needed. We are planning to
include metals in the near future. More details about these models will
be found in Green, Fontaine, \& Chayer (in preparation).

We used both grids of synthetic optical spectra to analyze our two
spectra of PG 1219+534. Using a procedure very similar to that prescribed
by Saffer et al. (1994), we performed a simultaneous fit of the
available Balmer lines and helium lines within a given grid of profiles
from our recent model atmosphere calculations. For the NLTE grid,
this procedure led to the solution for the MMT spectrum shown in Figure
\ref{cap:spectro1}, with errors given on the parameters corresponding to
formal errors of the fit. More realistically, based on spectra obtained
on different nights to estimate external errors, we adopted $\Teff=$
33,600 $\pm$ 370 K, $\log g=5.810\pm0.046$, and $\log N({\rm He})/N({\rm
  H})=-1.49\pm0.08$. Figure 1 indicates that we get a quite decent fit
to all of the available Balmer and helium lines, except for the HeII
4686 line which shows a flux deficit compared to our best-fit
model. This confirms the finding of Heber et al. (2000) and points to a
shortcoming in current models. Maybe the inclusion of metals in future
models could cure this problem. In any case, since ours is a global fit,
and since the relatively weak HeII 4686 line plays only a minor role in the
overall picture, we like to think that ours are realistic estimates of
the atmospheric parameters of PG 1219+534. We note in this context that
our LTE solution leads to a fit of comparable quality (not shown here)
but with slightly different parameters: $\Teff=$ 33,120 $\pm$ 110 K,
$\log g=5.80\pm0.02$, and $\log N({\rm He})/N({\rm H})=-1.49\pm0.02$
(formal errors). This indicates that NLTE effects on the populations of
hydrogen and helium ions are small in the atmosphere of PG 1219+534.
Of course, of the two solutions, the NLTE one is to be preferred
because the physics is more accurately treated in that
case.

\begin{table*}

\begin{center}
\caption{CFHT High Time Resolution Photometric Observations of PG 1219+534.\label{cap:obslog}}

\begin{tabular}{cccccccc}
\hline
Night&
Run&
Date&
Start Time&
Start Time&
Sampling&
Total Number of&
Run length\tabularnewline
\#&
Number&
(UT)&
(UT)&
BJD 2453073.5 + &
Time (s)&
Data Points&
(hours)\tabularnewline
\hline
1&
cfh-104&
2004-03-09&
06:45&
0.292811 &
10&
2221&
6.17\tabularnewline
2&
cfh-105a&
2004-03-10&
06:19&
1.269846&
10&
1336&
3.71\tabularnewline
2&
cfh-105b&
2004-03-10&
10:02&
1.430144&
10&
1471&
4.09\tabularnewline
3&
cfh-106a&
2004-03-11&
07:21&
2.312399&
10&
1610&
4.47\tabularnewline
3&
cfh106b&
2004-03-11&
11:49&
2.500243&
10&
553&
1.54\tabularnewline
4&
cfh-107&
2004-03-12&
06:14&
3.266342&
10&
3369&
9.36\tabularnewline
\hline
\end{tabular}
\end{center}
\end{table*}

Figure \ref{cap:spectro2} shows our NLTE solution for the low-resolution spectrum, with
the quoted uncertainties there again referring to the formal errors of
the fit. More realistic estimates are at least double those, and we may
adopt $\Teff=$ 34040 $\pm$ 220 K, $\log g=5.86\pm0.06$, and $\log N({\rm
  He})/N({\rm H})=-1.51\pm0.06$. Within the quoted uncertainties, both
independent measurements are in excellent agreement. This is very
encouraging and alleviates some of the fear (see, e.g., Saffer et
al. 1994) that the use of only three Balmer lines (in the more limited
spectral range of our MMT observations) could introduce some systematic
effects. On the basis of a substantial sample of sdB stars observed both
with the MMT and with the 2.3 m Steward telescope (the spectra used here
are part of that program), we will establish elsewhere that there are
indeed no significant systematic differences between the two
approaches. Another extremely encouraging development is that, using
completely independent tools, the referee of this article, Dr. Uli
Heber, has kindly communicated to us that his fit to our low-resolution
spectrum using comparable models to ours (i.e., NLTE H/He models with no
metals) leads to $\Teff=$ 34200 $\pm$ 70 K, $\log g=5.86\pm0.01$, and
$\log N({\rm He})/N({\rm H})=-1.45\pm0.02$ (formal errors), in
excellent agreement with our results. This gives added confidence that
our estimates of the atmospheric parameters of PG 1219+534 are secure.

In practice, in what follows, we will adopt our NLTE solution for the
MMT spectrum as our best spectroscopic estimates of the surface
parameters of \object{PG 1219+534}. To repeat, those are $\Teff=$
33,600 $\pm$ 370 K, $\log g=5.810\pm0.046$, and $\log N({\rm He})/N({\rm
  H})=-1.49\pm0.08$. These values are consistent with those derived by
\citet{1999MNRAS.305...28K}. Compared to the \citet{2000A&A...363..198H}
results, our values indicate a surface gravity somewhat lower than
proposed by them. Nonetheless, we point out that a much closer agreement
exists with their ``NLTE: H+He'' measurement (see their Table 5). In
fact, for the reasons given above concerning the question of mode
density, we may consider that measurement as ``their'' best solution as
well. Finally, we stress that the
derived atmospheric parameters are typical of most EC14026 stars and
place \object{PG 1219+534} close to the center of the theoretical
instability strip discussed by \citet{2001PASP..113..775C}.

\subsection{Time Series Photometry}

\begin{figure*}
\begin{center}\includegraphics[%
  scale=0.44]{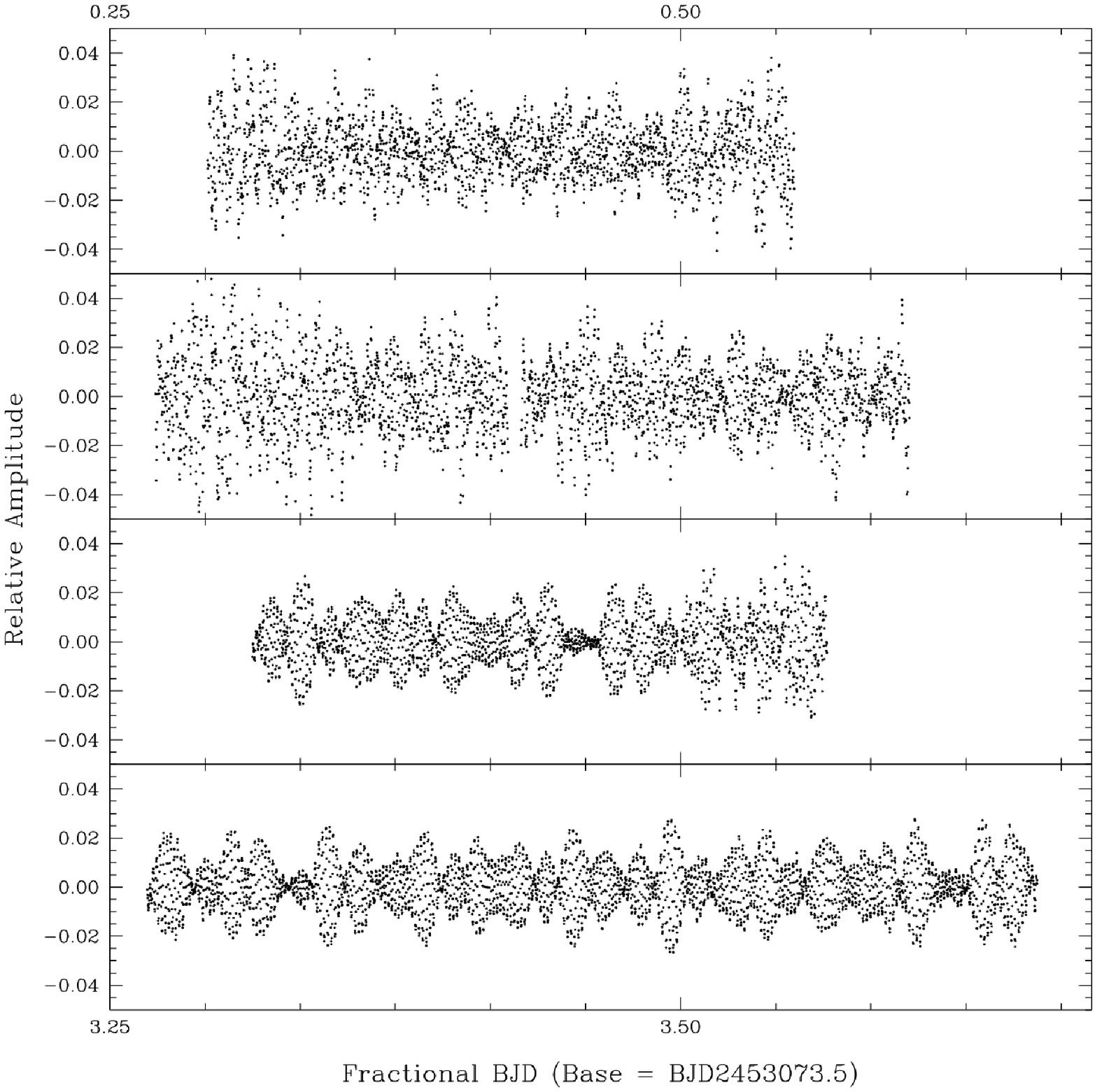}
  \hspace{5mm}
  \includegraphics[%
  scale=0.44]{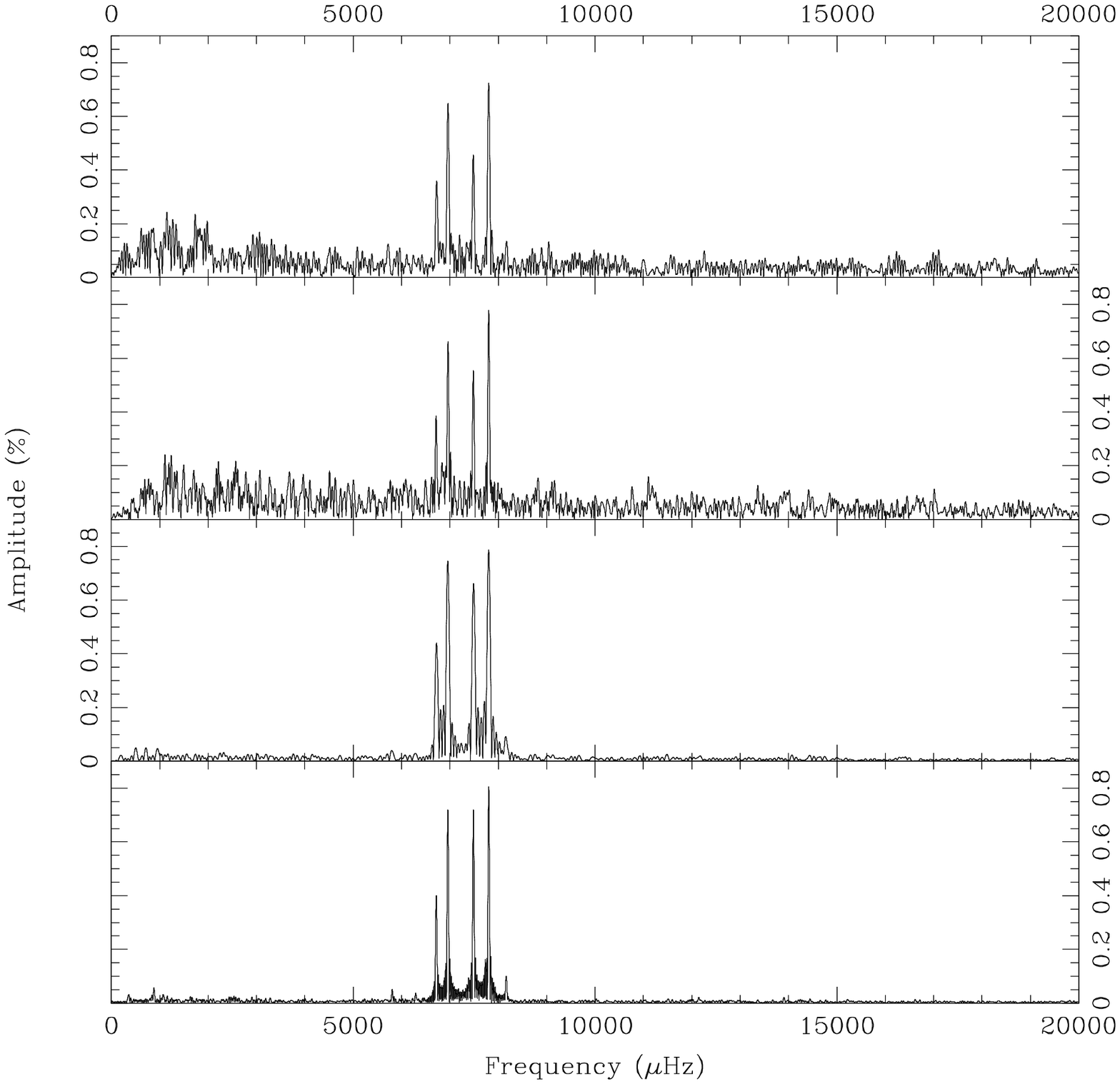}\end{center}

\caption{\emph{Left panel}: Optical (``white light'') light curves of
PG~1219+534 obtained at CFHT between 2004, March $9-12$ (\emph{top
to bottom slices}). The light curves are expressed in terms of residual
amplitude relative to the mean brightness of the star. \emph{Right
panel}: Fourier amplitude spectra (given as a percentage of the mean
brightness of the star) in the $0-20$ mHz bandpass associated with
each light curve shown in the \emph{left panel}.\label{global-lc}}
\end{figure*}

We observed \object{PG~1219+534} in optical ``white light'' fast
photometric mode at the CFHT during a dedicated four night run scheduled
on 2004, March. Table \ref{cap:obslog} provides a journal of these
observations. As with our previous programs, the photometric data
were gathered with \noun{Lapoune}, the portable Montr\'eal three-channel
photometer. The data were then carefully corrected for various
undesirable effects (e.g., atmospheric transparency variations,
extinction, etc.) using the data reduction package \noun{Oscar}
specifically designed by one of us (P.B.) for that purpose.
We refer the interested reader to \citet{1997ApJ...487L..81B}
for details on the observations and data reduction procedures.

\begin{figure}
\begin{center}
\hfill\vspace{2mm}\\
\includegraphics[scale=0.41]{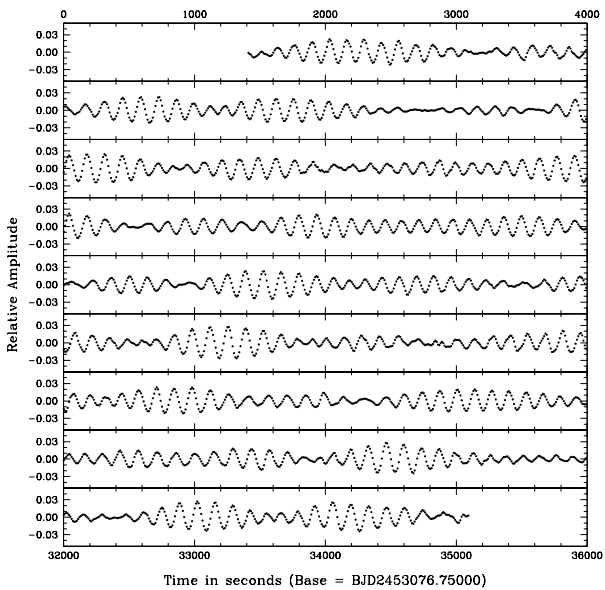}\end{center}

\caption{Expanded view of the best optical light curve obtained on PG~1219+534
during the night of 2004, March 12. Each band covers an interval of
4,000 s and the amplitude is expressed in terms of the residual relative
to the mean brightness of the star.\label{night4-lc} }
\end{figure}

The observing conditions varied from relatively poor for the first two nights
to simply outstanding for the rest of the run. The quality of the
data gathered during these observations is illustrated on a night-by-night
basis in Figure \ref{global-lc}, where the \emph{left panel} shows
the fully reduced light curves, and the \emph{right panel} displays
the corresponding Fourier amplitude spectra in the relevant $0-20$
mHz frequency bandpass. The light curves from the first and second
nights (runs cfh-104, cfh-105a and cfh-105b), as well as from the
very end of the third night (run cfh-106b) were rather noisy, and
this is also clearly reflected in their associated Fourier amplitude spectra.
These time series were obtained during bright time (contrary to our
request) and the presence of dust and high winds as well as thin cirrus
during the first two nights is responsible for the much lower quality of
the data we obtained. Our past experience at the CFHT shows that photometric
data gathered with \noun{Lapoune} can be quite good even in presence
of thin cirrus, but this works only under dark sky conditions. The
contrast is particularly sharp when comparing these data with the very
high quality photometry (more typical of what we have been accustomed to
at CFHT) acquired during most of the third night (run cfh-106a) and
during the fourth night (run cfh-107), where conditions became
photometric. That last light curve in particular, a chunk of 9.36 h with
no interruption obtained during perfectly photometric conditions, is
among the best that have been obtained with \noun{Lapoune} at CFHT, so far.
It is shown in greater detail in Figure \ref{night4-lc}. The variations
have typical peak-to-peak amplitudes of less than $\sim 6\%$ and are
dominated by a pseudo-period of $\sim130$ s. It is obvious however
that several other periodicities are involved in the variations, considering
the strong beating structures that can be seen in the light curve
and which are typical of destructive and constructive interferences
between modes. The multiperiodic nature of the variations is, of
course, confirmed in the Fourier amplitude spectra shown in the \emph{right
panel} of Figure \ref{global-lc}, where four dominant peaks clearly
emerge over the mean noise level. Our thorough analysis of the light
curves described in the following section will reveal that even more
periodicities, indeed, contribute to the luminosity variations seen
in \object{PG~1219+534}.

\section{Analysis of the Light Curves}

\begin{figure*}
\begin{center}\includegraphics[%
  scale=0.44]{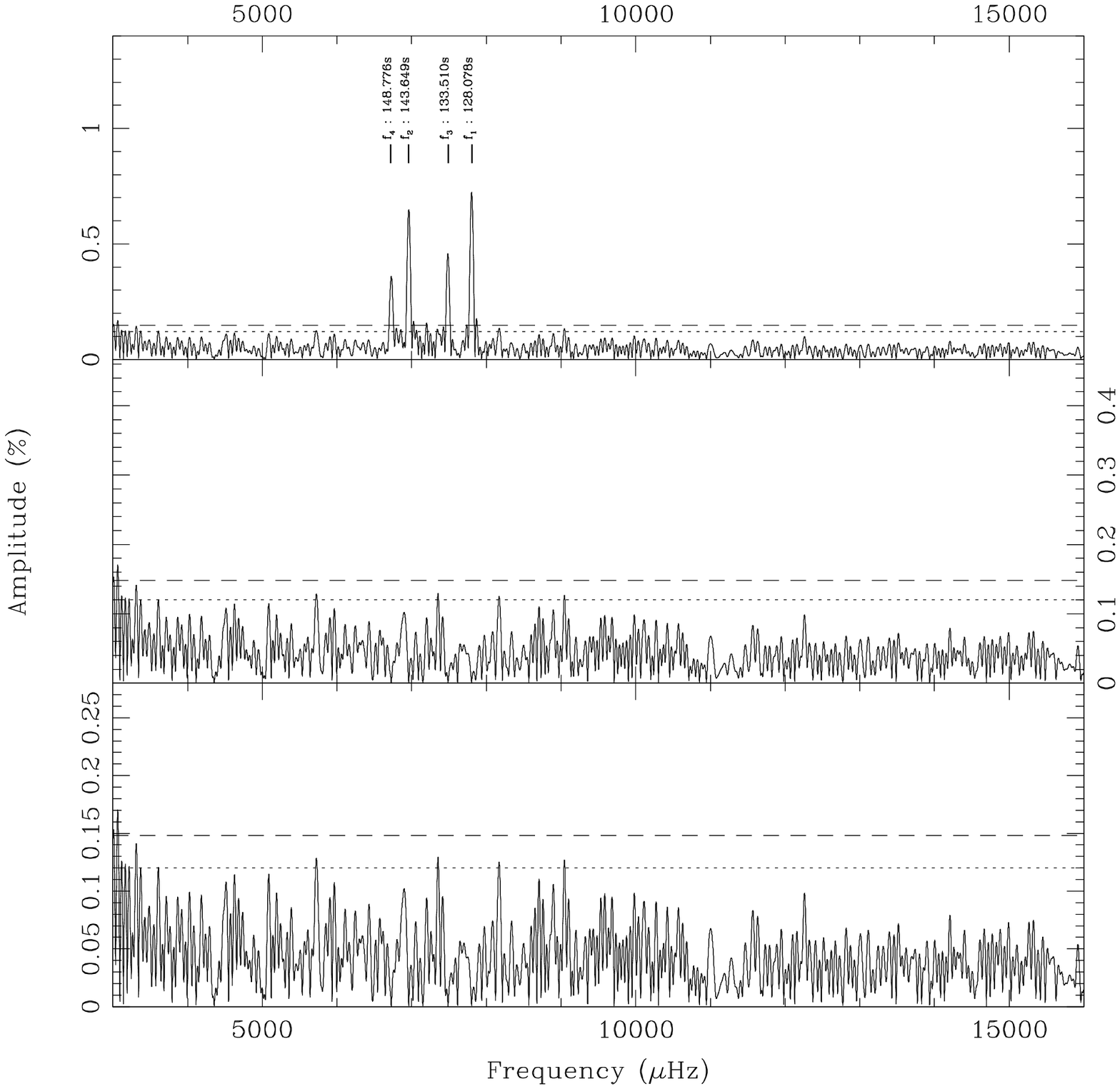}
  \hspace{5mm}
  \includegraphics[%
  scale=0.44]{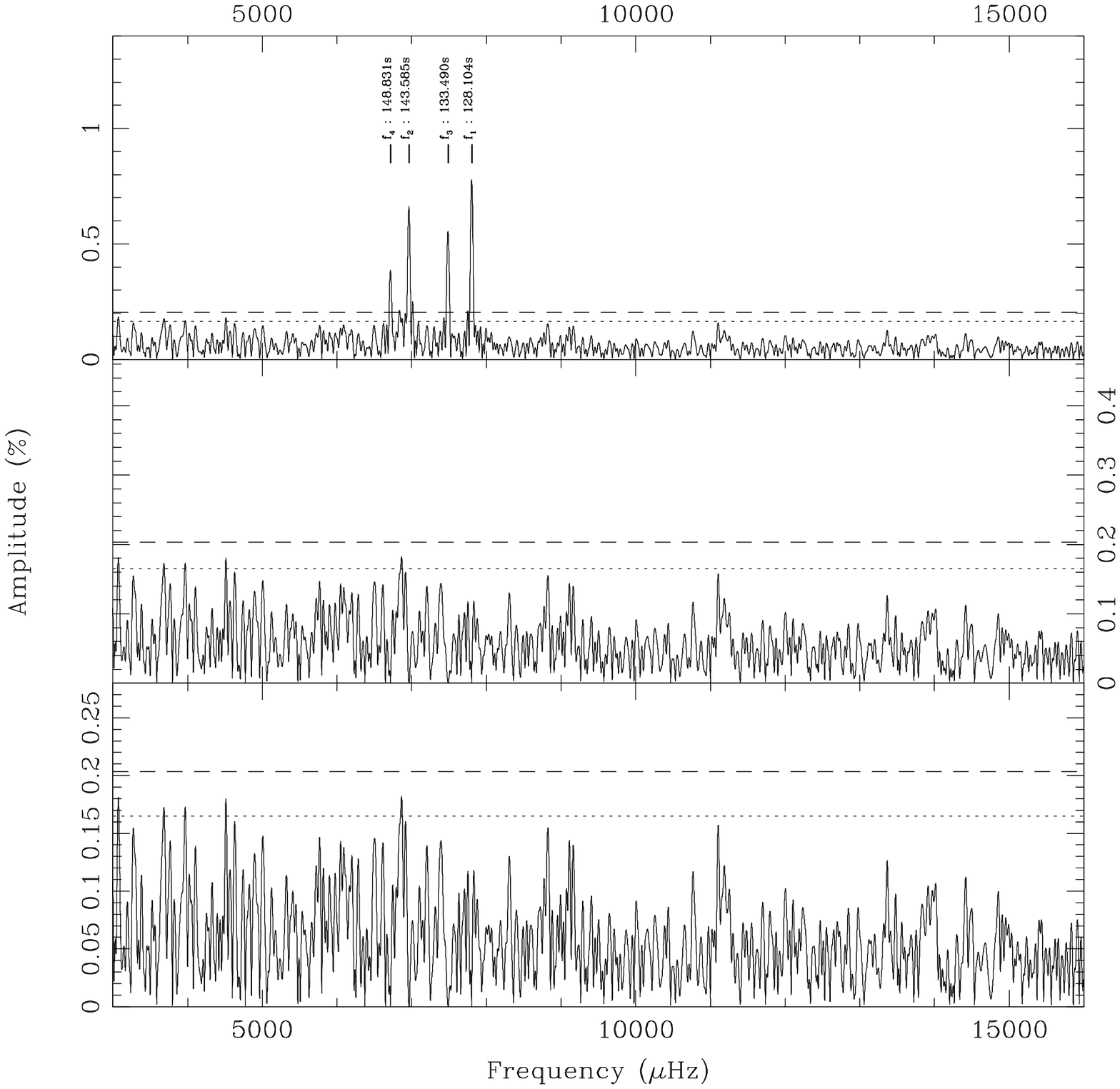}\end{center}

\begin{center}\includegraphics[%
  scale=0.44]{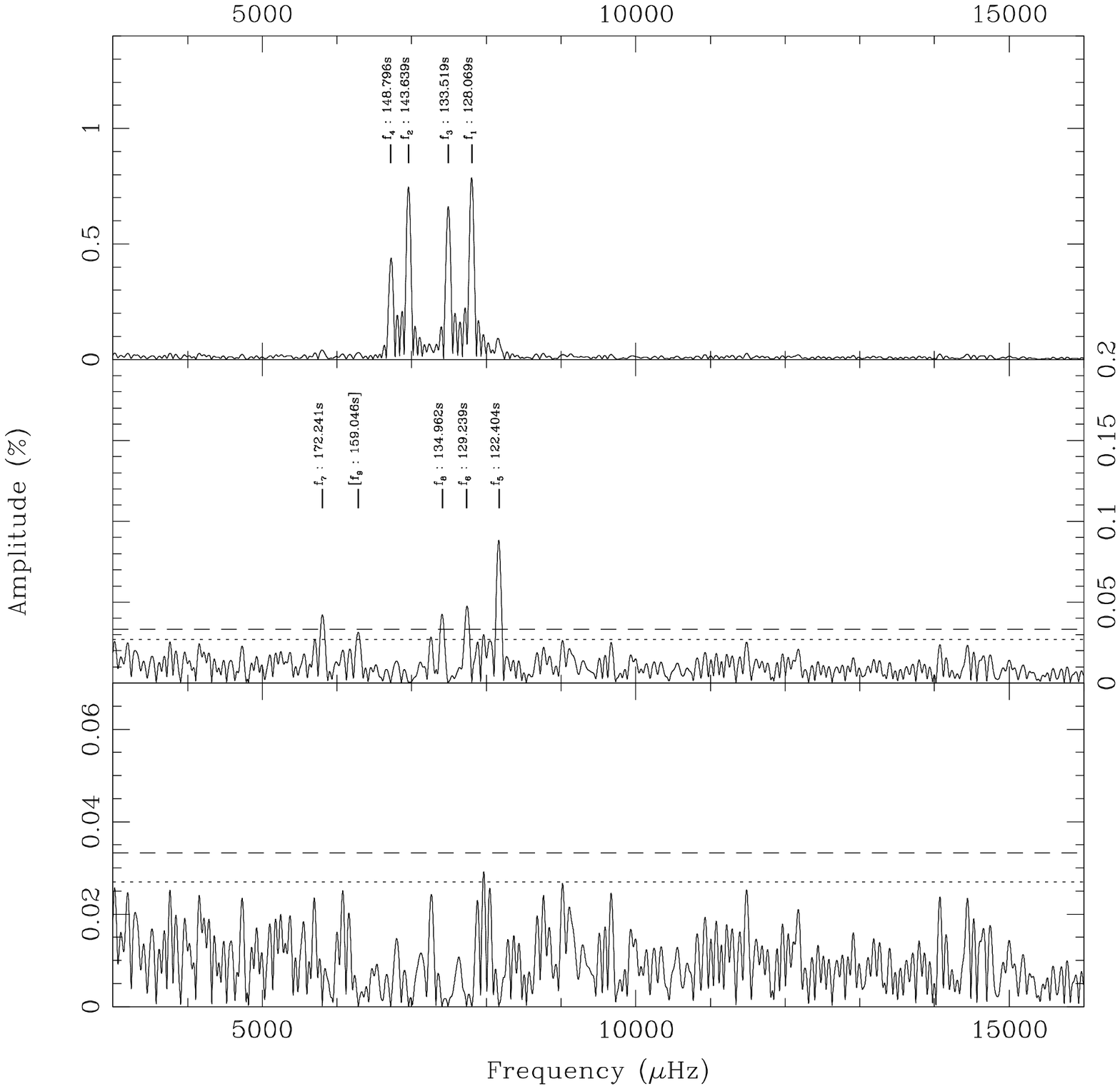}
  \hspace{5mm}
  \includegraphics[%
  scale=0.44]{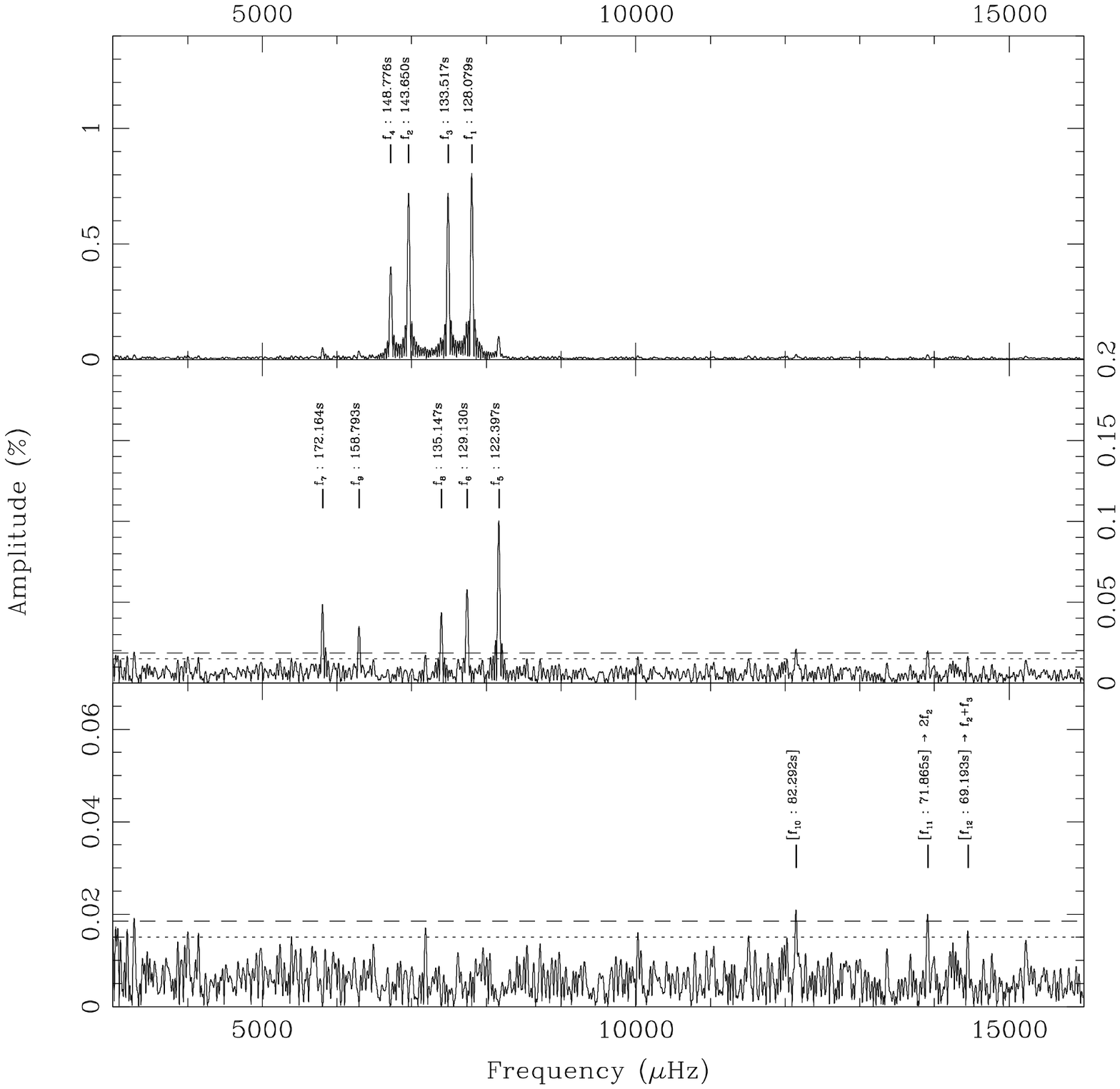}\end{center}

\caption{Fourier amplitude spectrum and residuals after prewhitening of the
identified periods (frequencies) in the time series of the four nights
analyzed separately (night 1, 2, 3 and 4, from \emph{upper-left panel}
to \emph{lower-right panel,} respectively). For night 3 (\emph{lower-left}
panel), only the best part (run cfh-106a) of the light curve has been
considered\label{prewhite1}. The dashed (dotted) line refers to a value
equal to 3.7 (3.0) times the mean noise level.}
\end{figure*}

Considering that the quality of the light curves obtained on \object{PG~1219+534}
during our CFHT run is particularly inhomogeneous, we carried out
detailed frequency analyses on various subsets of the observations.
We analyzed the time series in a standard manner by combining Fourier
analysis, least-squares fits to the light curves, and prewhitening
techniques. The procedure to extract pulsation modes from the light
curves is now greatly eased by a new dedicated software named \noun{Berthe,}
developed recently by one of us (P.B.). This software provides efficient
tools to conveniently proceed along the procedure described below.

\begin{figure*}
\begin{center}\includegraphics[%
  scale=0.44]{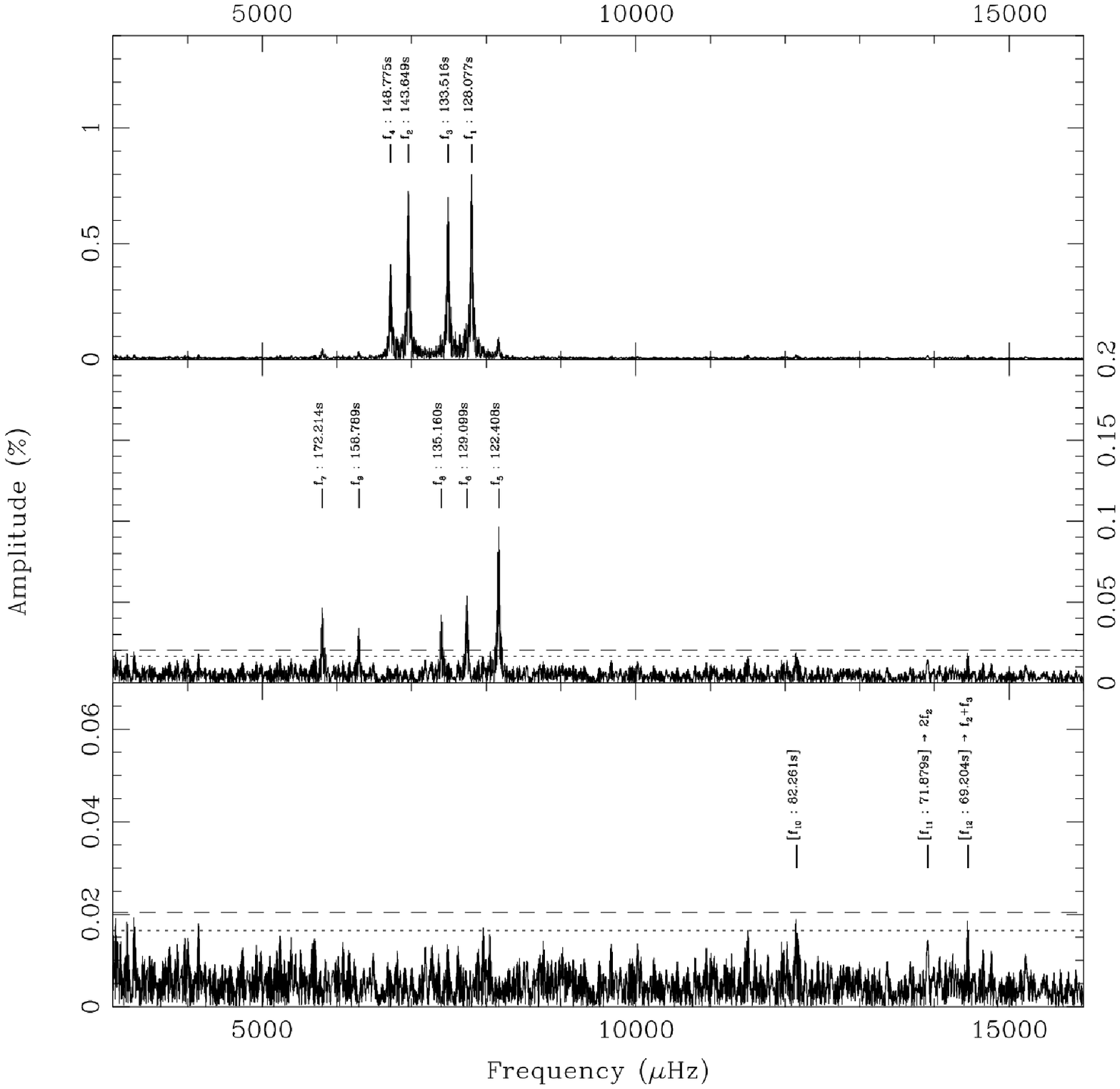}
  \hspace{5mm}
  \includegraphics[%
  scale=0.44]{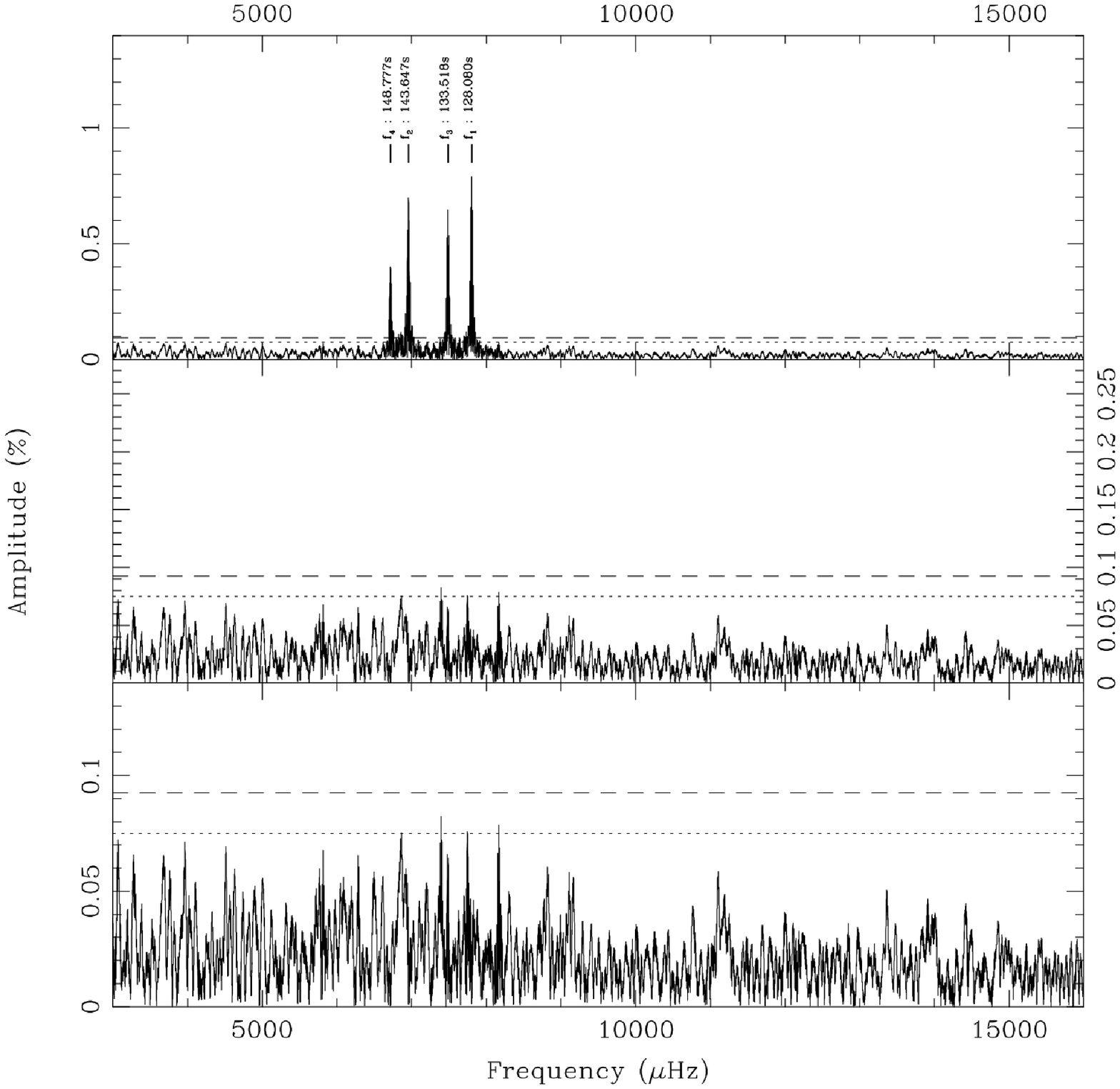}\end{center}

\caption{Same as Figure \ref{prewhite1}, but for the combined time series
from night 3 (run cfh-106a only) and night 4 (\emph{left panel}),
and for all light curves assembled together (\emph{right panel}).\label{prewhite2}}
\end{figure*}

The first step of the procedure is to compute the Fourier transform
of the light curve. We typically use 10,000 frequency points per chunk
of width 1 mHz in the Fourier space, which largely oversamples that
domain. The upper slices in the panels shown in Figure \ref{prewhite1}
and Figure \ref{prewhite2} illustrate the Fourier spectrum of the
light curves of \object{PG~1219+534} in the relevant 3-16 mHz bandpass.
As noted previously, the light curve shows multiperiodic harmonic
oscillations which is reflected through the presence of multiple peaks
in the Fourier spectrum. We next isolate from the Fourier spectrum
the frequency (period) of the usually (but not always) largest peak
seen in a given frequency interval. That frequency (period) is used
in a least-squares routine that provides the amplitude and the phase
of the harmonic oscillation of that given period which best-fits the
complete light curve. This harmonic oscillation is then subtracted
from the light curve in the time domain (prewhitening) and the Fourier
transform of the residual light curve is computed. A second frequency,
again often associated with the highest peak in the Fourier transform
of the residual, is identified and the least squares routine is reapplied
but this time with the two frequencies, amplitudes, and phases fitted
\emph{simultaneously}. This provides upgraded estimates for the amplitude
and phase of the first harmonic oscillation and initial estimates
for the second one. These two harmonic oscillations are then subtracted
from the original light curve through prewhitening and another Fourier
transform of the residual light curve is computed. The procedure is
repeated until no useful information can be extracted from the residual
light curve through this method. It is standard to stop the iteration
when the amplitude of the candidate peaks in the Fourier domain gets
below 3.7 times the local rms noise level, which corresponds to the
99\% statistical significance limit. We find, in practice, that applications
of more sophisticated statistical methods to test for the reality
of peaks in the Fourier domain (such as the false-alarm probability
formalism) generally confirm that an amplitude 3 times larger than
the mean noise level is sufficient to account for the reality of a
mode.

\begin{table*}

\caption{Periods of the Modulations Detected in the Light Curves of PG~1219+534.\label{cap:Harmonic-Oscillations-Detected}}

\begin{center}\begin{tabular}{lcccccccc}
\hline
&
Koen \emph{et al}.&
Night&
Night&
Night&
Night&
Nights&
Nights&
Nights\tabularnewline
&
1999&
1&
2&
3&
4&
1 \& 2&
3 \& 4&
All\tabularnewline
\hline
Resolution ($\mu$Hz)&
3.64&
45.02&
35.03&
62.11&
29.68&
8.85&
8.61&
3.44\tabularnewline
Duty cycle (\%)&
17.8&
100&
98.3&
100&
100&
44.5&
42.9&
34.4\tabularnewline
FT Noise (\%)&
0.040&
0.040&
0.055&
0.0090&
0.0040&
0.040&
0.0055&
0.025\tabularnewline
\hline
\hline
&
Period&
Period&
Period&
Period&
Period&
Period&
Period&
Period\tabularnewline
Id.&
(s)&
(s)&
(s)&
(s)&
(s)&
(s)&
(s)&
(s)\tabularnewline
\hline
$f_{7}$&
...&
...&
...&
172.241&
172.164&
...&
172.214&
...\tabularnewline
$f_{9}$&
...&
...&
...&
159.046&
158.793&
...&
158.789&
...\tabularnewline
$f_{4}$&
148.776&
148.603&
148.831&
148.796&
148.776&
148.769&
148.775&
148.777\tabularnewline
$f_{2}$&
143.649&
143.619&
143.585&
143.639&
143.650&
143.652&
143.649&
143.647\tabularnewline
$f_{8}$&
{[}134.960{]}&
...&
...&
134.962&
135.147&
...&
135.160&
...\tabularnewline
$f_{3}$&
133.510&
133.497&
133.490&
133.519&
133.517&
133.503&
133.516&
133.518\tabularnewline
$f_{6}$&
...&
...&
...&
129.239&
129.130&
...&
129.099&
...\tabularnewline
$f_{1}$&
128.078&
128.115&
128.104&
128.069&
128.079&
128.075&
128.077&
128.080\tabularnewline
$f_{5}$&
...&
...&
...&
122.404&
122.397&
...&
122.408&
...\tabularnewline
$[f_{10}]$&
...&
...&
...&
...&
{[}82.292{]}&
...&
{[}82.261{]}&
...\tabularnewline
$[f_{11}]$&
...&
...&
...&
...&
{[}71.865{]}&
...&
{[}71.879{]}&
...\tabularnewline
$[f_{12}]$&
...&
...&
...&
...&
{[}69.193{]}&
...&
{[}69.204{]}&
...\tabularnewline
\hline
\end{tabular}\end{center}
\end{table*}

We first analyzed the light curves obtained on \object{PG~1219+534}
separately, on a night-by-night basis, before combining them according
to their respective mean noise level, i.e., night 1 was combined with
night 2, and night 3 (without the noisy chunk at the end of the night
that would just ruin the signal-to-noise ratio, if added)
was combined with night 4. Finally, we examined the complete light
curve built from all nights. For the various cases considered, illustrations
of the Fourier transforms obtained before and after prewhitening of
the identified frequencies are provided in Figure \ref{prewhite1}
and Figure \ref{prewhite2}. The periods extracted from these analyses
are listed in Table \ref{cap:Harmonic-Oscillations-Detected} (each
identified period is labeled $f_{j}$, where $j$ is ordered by decreasing
amplitude), as well as the results obtained previously by
\citet{1999MNRAS.305...28K}.
In this table, we also provide information relative to the run considered.
This includes the resolution achieved (in $\mu$Hz), the rms noise
level reached in the data (in $\%$ of the mean brightness of the
star), and the duty cycle (in \% of the total coverage). Of course,
each night considered individually gives a duty cycle of $100\%$
(except for the second night, due to a short interruption in the time
series), which then decreases due to the daily gaps that are introduced
when data sets of successive nights are combined (down to 34.4\% for
the whole run). In the meantime, however, the frequency resolution
improves from 29.7 $\mu$Hz for our longest contiguous light curve
(night 4) down to 3.44 $\mu$Hz with the time baseline of the whole
photometric run. We note that the best frequency resolution
reached with our data set is comparable to the resolution obtained
in \citet{1999MNRAS.305...28K}, but our coverage is better. Unfortunately,
no significant improvement of the mean noise level in the Fourier
space can be achieved with the whole data set compared to
\citet{1999MNRAS.305...28K} data. This is a direct consequence of the
heterogeneous quality of the CFHT light curves. The rms noise level in
the Fourier domain is approximately $5-10$ times higher for the first
two nights compared to the excellent data obtained during nights 3 and
4. Mixing all time series to construct the global light curve
significantly degrades the S/N ratio in the Fourier domain, as the mean noise
level is essentially regulated by the light curve of worst quality.
Obviously, in the present case, focussing on the best data only (i.e.,
those obtained during night 3 and night 4) would yield the best results for
extracting pulsation modes from \object{PG~1219+534} photometric
variations.

The light curve from the fourth night is by far the best obtained
during these observations and thus, not surprisingly, gives the Fourier
transform with the lowest rms noise level ($\sim0.004\%$). From that
time series alone, 9 frequencies (periods) could be extracted without
any ambiguity. The prewhitening sequence for this particular light curve
is illustrated in the successive bands shown in the lower-right panel
of Figure \ref{prewhite1}. The upper band displays the full amplitude
spectrum in the relevant $3-16$ mHz frequency bandpass where photometric
activity is concentrated. We note that from 16 mHz up to the Nyquist
frequency (50 mHz, as we used a sampling time of 10 s), the structures
found in the Fourier amplitude spectrum are entirely consistent with
noise. At the low frequency end, the rms noise level increases due
to residual atmospheric variations (which are difficult to remove
completely) and no obvious periodicities could be identified (but
see below). Four dominant periods, labeled $f_{1}$ to $f_{4}$, are
clearly seen at that stage, while the noise level is so low at the
provided scale that it is essentially invisible. The mid-band shows
the Fourier transform of the residual light curve after prewhitening
of those four main harmonic oscillations. This leads to the unambiguous
detection of five additional periods ($f_{5}$ to $f_{9}$) that clearly emerge
above the detection threshold of 3.7 times the rms noise level materialized
by the horizontal dashed line (the dotted line indicates the limit
of 3 times the rms noise level). The lower band displays the Fourier
transform of the residual light curve after prewhitening of the nine
identified periodicities. Little power is then left in the Fourier
transform of the residual light curve, but hints of remaining structures
exist, in particular at the high-frequency end of the bandpass considered.
Three more structures ($f_{10}$, $f_{11}$, and $f_{12}$) are indeed
found slightly above, or just below the detection threshold (dashed line).
These are also visible in the Fourier transform of the residual light
curve of the combined nights 3 and 4 (lower-band in left-panel of
Figure \ref{prewhite2}), but unfortunately cannot be confirmed with
the analysis of the other light curves, due to their much higher mean
noise level. For this reason, we consider as marginal the detections
of these periods, although we will provide below arguments which suggest
that these structures may be real. These marginal period detections
are indicated within brackets in Table \ref{cap:Harmonic-Oscillations-Detected}.

\begin{table*}

\caption{Properties of the Harmonic Oscillations Identified in \object{PG~1219+534}.\label{cap:Identified-frequencies-for}}

\begin{center}\begin{tabular}{lccccl}
\hline
&
Frequency&
Period&
Amplitude&
Phase&
Comments\tabularnewline
Id.&
($\mu$Hz)&
(s)&
(\%)&
(s)&
\tabularnewline
\hline
$f_{7}$&
5806.71&
172.214&
$0.0456\pm0.0030$&
$68.72\pm1.80$&
\tabularnewline
$f_{9}$&
6297.65&
158.789&
$0.0334\pm0.0030$&
$23.14\pm2.27$&
\tabularnewline
$f_{4}$&
6721.55&
148.775&
$0.4053\pm0.0030$&
$42.86\pm0.18$&
detected with an amplitude of 0.22\% by \citet{1999MNRAS.305...28K}
\tabularnewline
$f_{2}$&
6961.39&
143.649&
$0.7228\pm0.0030$&
$102.50\pm0.09$&
detected with an amplitude of 0.18\% by \citet{1999MNRAS.305...28K}
\tabularnewline
$f_{8}$&
7398.62&
135.160&
$0.0429\pm0.0030$&
$98.73\pm1.51$&
\tabularnewline
$f_{3}$&
7489.73&
133.516&
$0.6874\pm0.0030$&
$43.65\pm0.09$&
detected with an amplitude of 0.23\% by \citet{1999MNRAS.305...28K}
\tabularnewline
$f_{6}$&
7745.98&
129.099&
$0.0544\pm0.0030$&
$104.69\pm1.13$&
\tabularnewline
$f_{1}$&
7807.79&
128.077&
$0.7972\pm0.0030$&
$44.07\pm0.08$ &
detected with an amplitude of 0.85\% by \citet{1999MNRAS.305...28K}
\tabularnewline
$f_{5}$&
8169.43&
122.408&
$0.0963\pm0.0030$&
$95.70\pm0.61$&
\tabularnewline
$[f_{10}]$&
{[}12156.40{]}&
{[}82.261{]}&
{[}$0.0190\pm0.0030${]}&
{[}$59.38\pm2.07${]}&
slightly above detection threshold; not used for seismology\tabularnewline
$[f_{11}]$&
{[}13912.31{]}&
{[}71.879{]}&
{[}$0.0143\pm0.0030${]}&
{[}$21.06\pm2.40${]}&
$2f_{2}$; not used for seismology\tabularnewline
$[f_{12}]$&
{[}14450.08{]}&
{[}69.204{]}&
{[}$0.0188\pm0.0030${]}&
{[}$48.10\pm1.76${]}&
$f_{2}+f_{3}$; not used for seismology\tabularnewline
\hline
\end{tabular}\end{center}
\end{table*}

Globally, we find that the four dominant modes, $f_{1}$ to $f_{4}$,
are detected in all the CFHT light curves (and combinations of light
curves) and correspond to the four periods already identified in \citet{1999MNRAS.305...28K}.
These authors had also suspicions concerning a fifth period at 134.96
s (that corresponds to $f_{8}$, here), which we indeed confirm with
no ambiguity on the basis of the best light curves we obtained. Note
that only the four dominant periods are seen when nights 1 and 2 are
considered either separately or combined into larger data sets, such
as the complete light curve built from all nights. This is due, again,
to the relatively poor quality of these specific data and the dramatic
impact it has on the mean noise level in the Fourier domain. Nonetheless,
the analysis of the best available time series yields a significant
improvement in terms of the number of modes detected compared to the
original data, since five additional periods ($f_{5}$ to $f_{9}$)
can be securely identified in the spectrum of \object{PG~1219+534}.

We provide in Table \ref{cap:Identified-frequencies-for} a list of
the detailed properties that characterize the nine harmonic oscillations
uncovered. These are the frequency (in $\mu$Hz), the period (in second),
the amplitude (in \% of the mean brightness), and the phase (in seconds).
The three marginal detections mentioned above are also included within
brackets. We note that two of these can be interpreted as harmonics
or cross-frequencies of the dominant peaks, i.e., $f_{11}=2f_{2}$
and $f_{12}=f_{2}+f_{3}$ (and therefore would not constitute independent
modes usable for asteroseismology), while the third period $(f_{10}$)
may be an independent pulsation mode. We have, however, explicitly
excluded this period and considered only the nine well secured pulsation
modes for the following asteroseismic analysis of \object{PG~1219+534}.
We have adopted the list of periods derived from the analysis of the
combined light curves from night 3 and night 4, which allows to recover
all the harmonic oscillations seen in the data taken during night 4
treated separately (the best achievement in terms of S/N ratio) while
reaching a decent frequency resolution of 8.61 $\mu$Hz. The measured
frequencies (each associated with a frequency peak in the Fourier
domain) are probably accurate to 1/10 of the formal resolution (itself
corresponding to the width at half-maximum of such a peak), i.e., to
within 0.86 $\mu$Hz (see \citealp{1988bretthorst.book.....V}).
The amplitudes and phases derived through the
least-squares technique have formal errors as given in Table
\ref{cap:Identified-frequencies-for}. We note at this point, that the
gain in frequency resolution obtained while combining all light curves
for the Fourier analysis does not reveal the presence of close frequencies
near the dominant peaks (due to rotational splitting, for
instance). This observation holds for the five newly discovered
frequencies as well, although the frequency resolution is lower in these
cases. Such a finding indicates that \object{PG~1219+534} is likely a
slow rotator with a rotation period longer than $\sim3.4$ days, i.e.,
the longest time baseline achieved with the photometry currently
available. Finally, we note a significant change in the amplitude of
three of the four dominant periods that have been detected in both our
contemporary data and the older \citet{1999MNRAS.305...28K} light curves.
Except for the period $f_1$ which is found slightly weaker (by $\sim 6\%$) in
our time series, the periods $f_2$, $f_3$, and $f_4$ all appear much stronger than
observed in \citet{1999MNRAS.305...28K}, by factors up to $\sim 2- 4$. Such
amplitude variations are not uncommon among rapidly pulsating sdB stars.
In \object{PG~1219+534}, these may be due to beating between very close
-- thus, unresolved with current data -- frequencies (multiplet components
caused by slow rotation, for instance), and/or to significant changes in the intrinsic
amplitudes of the modes themselves. However, additional observations with longer
time baselines will be needed to further discuss the stability of
pulsation mode amplitudes in this star.

Interestingly, the odd conditions under which the CFHT data were obtained
provides us with a perfect illustration of the fact that mixing
low and high S/N ratio time series can essentially ruin all
the benefit of having the high-sensitivity data. The overall mean
noise level in the Fourier space is clearly mostly regulated by the
noise level of the worst light curve included in the analysis. It
is instructive that, for this particular star \object{PG~1219+534}
which, at the outset, turns out to have a relatively simple and well
resolved pulsation spectrum, just one night of high sensitivity CFHT
data appears to be sufficient to make this object amenable to detailed
asteroseismic analysis. More generally, for the purpose of asteroseismology
of stars with simple enough oscillation spectra (such as EC14026 stars
often have, with some exceptions), high sensitivity is often to
be sought preferrentially to frequency resolution and/or
coverage. This, indeed, has some implications for the treatment of
multisite campaign data where light curves of various quality are
generally mixed. In those cases, mixing low S/N data with
high S/N data, although improving resolution and reducing aliasing
problems, can significantly degrade the S/N ratio reached with the
best available data. Accordingly, weighting techniques may
contribute, in some circumstances, to reduce the deterioration of
the overall S/N ratio, although generally at the expense
of a degraded window function (see, for instance, the discussion of
\citealt{Handler2003}). However, with very heterogeneous
data sets, these techniques would still lead to a significant increase of the
mean noise level in the Fourier domain compared to the S/N ratio achieved with
the best light curves.
This has to be considered carefully when the goal is to extract low amplitude
variations to increase the number of modes for asteroseismology.
Of course, the best is to have both high
sensitivity \emph{and} coverage with data as \emph{homogeneous} as
possible. Along this line, we hope for future bi- (or tri-) site fast
photometric observations based on 4m-class telescopes, as it would
constitute, in addition to other developing techniques (multicolor
photometry and high time resolution spectroscopy), a significant step
forward in the field of asteroseismology of EC14026 stars.

To end this Section, let us briefly mention that, while performing the
Fourier analysis of the best light curves obtained on \object{PG~1219+534},
we have discovered hints of low frequency modulations in the amplitude
spectrum. Having a closer look at the highest S/N ratio Fourier amplitude
spectra shown in the right panel of Figure \ref{global-lc} (especially
the lowest band corresponding to the data of our best night), one
can notice low amplitude structures below 1 mHz that still significantly
emerge above the local mean noise level. If such periodic modulations
are real and due to pulsations, those peaks would correspond to modulations
with periods much longer than the typical periods observed in EC 14026
stars, thus involving relatively high-order gravity modes. To some
extent, it would be similar to the variations seen in the long period
sdB pulsators, although \object{PG~1219+534} does not have, in principle,
the expected physical parameters to belong to that class of pulsating
stars which, typically, have much lower effective temperatures and
surface gravities. A possible analogy would be with the recent exciting
discovery of a long-period photometric variation in the EC14026 star HS
0702+6043 \citep{2005whdw.conf...press} which suggests that this star
may pulsate in both low-order $p$-modes and high-order $g$-modes.
It is, however, premature to speculate further on the nature of these
(potential) low frequency luminosity modulations in \object{PG~1219+534}.
Additional photometric observations at high sensitivity with a longer
time baseline than is currently available will be necessary to address
this particular issue and possibly confirm (or rule out) the presence of
$g$-modes in \object{PG~1219+534}. Specific efforts along this line are
underway.

\section{Asteroseismic Interpretation of the Observations}

\subsection{The Double-Optimization Scheme}

The asteroseismological interpretation of the period spectrum of pulsating
stars has often been impaired by the lack of identification of the
modes being observed. The analysis of EC14026 stars is confronted
to similar challenges as white light fast photometry generally gives
little clue -- and quite often no clue at all -- about the degree $\ell$
and/or radial order $k$ of the modes detected. Indeed, for \object{PG
  1219+534}, we have no \emph{a priori} knowledge of the geometry
of the nine observed modes (not even hints on the $\ell$ value, since
no splitting due to rotation has been detected at the current frequency
resolution). Nonetheless, we have pursued major efforts in the last
few years to develop a new method that allows us to bypass this difficulty.

Our approach is based on the well known forward method which consists
of comparing directly periods computed from stellar models to periods
observed in the star under consideration with the goal of reproducing
as accurately as possible its oscillation spectrum. Applied to EC14026
stars, it consists of a double-optimization procedure that takes place
simultaneously at the period matching level and in the model parameter
space. Due to the lack of mode identification from the available data,
assigning $N_{o}$ observed periods to $N_{t}$ theoretical periods
from a given model ($N_{t}>N_{o}$, in general) usually leaves a myriad
of possible combinations. The first optimization therefore consists
of finding the combination (or mode identification) that leads to
the best possible \emph{simultaneous} match of all the observed periods
for that given model. We note that this \emph{best} match does not
necessarily have to be a \emph{good} match at this point. This will
be the role of the second optimization procedure discussed below.
More formally, the quality (or merit) of a period fit is evaluated
as the sum of the squared differences $\chi^{2}$ between the observed
periods and their assigned theoretical periods (i.e., through a least-squares
formalism). The chosen mode identification for a given model is then
the combination that provides a $\overline{\chi}^{2}$ value corresponding
to the minimum of that sum among all the possible associations, that
is\begin{equation}
\overline{\chi}^{2}\equiv\min\left\{ \chi^{2}=\sum_{i=1}^{N_{o}}\left(\frac{P_{{\rm obs}}^{(i)}-P_{{\rm th}}^{(i)}}{\sigma_{i}}\right)^{2}\,,\,\forall\,\,{\rm combinations}\right\} .\label{eq:merit-function}\end{equation}
Here, $P_{{\rm obs}}^{(i)}$ is the $i^{{\rm th}}$ observed period
and $P_{{\rm th}}^{(i)}$ represents its associated theoretical period.
$\sigma_{i}$ is an optional weight that can be attributed either
individually on periods or globally (see Subsection 4.3). The second
optimization is carried out in the model parameter space where the
quantity $\overline{\chi}^{2}$ derived from the first optimization
is now seen as a function of the $N$ model parameters (considering
here the general case), i.e.,
$\overline{\chi}^{2}\equiv\overline{\chi}^{2}(a_{1},a_{2},...,a_{N})$,
where $a_{i}$ represents one of the $N$ parameters. The procedure
then consists of localizing the optimal model (or models) that minimizes
this {}``merit'' function in the $N$-dimensional space, leading
objectively to the best period match of the observations.

We have developed various sets of numerical tools that allows us to
apply this double-optimization scheme to isolate best period fitting
models for asteroseismology. Our original efforts along those lines
were described in \citet{2001ApJ...563.1013B}. Since then, many important
improvements have been made and two independent packages (fortunately
leading to the same results!) are now operational, one in Montr\'eal
and the other in Toulouse. Detailed descriptions of the numerical
tools developed in these packages will be reported elsewhere. Briefly,
the Toulouse package extensively used for this analysis of \object{PG~1219+534}
includes a genetic algorithm based period matching code that performs
the first optimization part, a parallel grid computing code useful
to explore the parameter space and visualize the complex shape of
the $\overline{\chi}^{2}(a_{1},a_{2},...,a_{N})$ function, and a
parallel multimodal genetic algorithm based optimization code designed
to explore efficiently the vast $N$-dimensional parameter space and
seek for the optimal solutions. By locating simultaneously the global
and local minima of the $\overline{\chi}^{2}$ function, this code
is capable of identifying several optimal models if more than one
solution exists.

It is important to realize that this approach is in fact built on
the very simple requirement that models pretending to provide a good
asteroseismological fit of the spectrum of a pulsating star must match
all the observed periods \emph{simultaneously}, i.e., the procedure
is a \emph{global} optimization. Moreover, we stress that the solutions
identified from this approach do not rely on any previous mode identification.
The mode identification appears instead naturally as the solution
of the global fitting procedure, i.e., the mode identification obtained
is the one that provides the best \emph{simultaneous} match of \emph{all}
the detected periods.

\subsection{Equilibrium Models and Computations of their Pulsation Properties}

Applications of our double-optimization scheme to EC14026 stars requires
computing appropriate theoretical period spectra that need to be
compared with the observed periods. Three codes are involved in this
process, starting with an equilibrium model building code first described
in \citet{1994esa..conf..560B} and adapted to produce the so-called
{}``second generation'' models suitable for pulsating sdB stars
(see \citealp{1997ApJ...483L.123C,2001PASP..113..775C}). These models
are improved static envelope structures extending as deep as $\log q=\log[1-M(r)/M_{*}]\simeq-0.05$.
For the purpose of pulsation calculations, the central missing nucleus
that contains $\sim10\%$ of the stellar mass is treated as a {}``hard
ball''. This approach, in comparisons to full evolutionary stellar
models, gives excellent pulsation results, especially for the relatively
shallow $p$-modes that are, in hot B subdwarf stars, insensitive
to the detailed structure of the central parts of the core as was
discussed at length in \citet{1999PhDT........26C} and
\citet{2000ApJS..131..223C,2002ApJS..139..487C}.
Moreover, these {}``second generation'' static models are superior
to all evolutionary structures currently available in that they incorporate
the nonuniform iron abundance profiles predicted by the theory of
microscopic diffusion assuming an equilibrium between gravitational
settling and radiative levitation. We recall that diffusion leads
to the constitution of a reservoir of iron in the H-rich envelope
of subdwarf B stars with significant iron enrichments produced locally
which are responsible, through the $\kappa$-mechanism, for the destabilization
of the low-order, low-degree $p$-modes observed in EC14026 pulsators
\citep{1997ApJ...483L.123C}. This additional ingredient in the constitutive
physics of the models is therefore essential to reproduce the excitation
of the pulsation modes. In addition, microscopic diffusion modifies
sufficiently the stellar structure to induce significant changes to
the pulsation periods themselves, thus impacting on the asteroseismic
analysis. Consequently, radiative levitation is a crucial ingredient
that \emph{must} be taken into account in the modeling of pulsating
sdB stars for the purpose of accurate asteroseismology. Four fundamental
parameters are required to specify the internal structure of hot B
subdwarf stars with the second-generation models: the effective temperature
$T_{{\rm eff}}$, the surface gravity $\log g$ (traditionally given
in terms of its logarithm), the total mass of the star $M_{*}$, and
the logarithmic fractional mass of the hydrogen-rich envelope $\log q({\rm H)\equiv}\log[M({\rm H})/M_{*}]$.
The latter parameter is intimately related to the more familiar parameter
$M_{{\rm env}}$, which corresponds to the total mass of the H-rich
envelope%
\footnote{We stress that the parameter $M_{{\rm env}}$ commonly used in Extreme
Horizontal Branch stellar evolution includes the mass of the hydrogen
contained in the thin He/H transition zone, while the parameter $M({\rm H})$
in our models does not. They are related through $\log[M_{{\rm env}}/M_{*}]=\log[M({\rm H})/M_{*}]+C$,
where $C$ is a small positive term, slightly dependent on the model
parameters, determined by the mass of hydrogen that is present inside
the He/H transition zone itself. %
}.

In a second step, we use two efficient codes based on finite element
techniques to compute the pulsation properties of the model. The first
one is an updated version of the adiabatic code described in detail
in \citet{1992ApJS...80..725B}. It is used as an intermediate (and
necessary) step to obtain estimates for the pulsation mode properties
(periods and eigenfunctions) that are then used as first guesses in
the solution of the full, nonadiabatic set of nonradial oscillation
equations. The second one is an improved version of the nonadiabatic
code that has been described briefly in \citet{1994ApJ...428L..61F}
and which solves the nonadiabatic eigenvalue problem. It provides
the necessary quantities to compare with the observations. For the
purpose of asteroseismology, these are mainly the periods and the
stability coefficients. To illustrate the type of quantities that
are derived from the whole process, we provide, in Table \ref{cap:Best-fit-model},
a typical output of the nonadiabatic pulsation code that gives the
pulsation properties of a given model. While it refers specifically
to the optimal model (see the discussion in the next subsection),
we only wish to emphasize the illustrative aspect of the theoretical
results at this point. For each mode found in the chosen period interval,
Table \ref{cap:Best-fit-model} gives the degree $\ell$, the radial
order $k$, the period $P_{{\rm th}}$ ($=2\pi/\sigma_{R}$, where
$\sigma_{R}$ is the real part of the complex eigenfrequency), the
stability coefficient $\sigma_{I}$ (the imaginary part of the eigenfrequency),
the logarithm of the so-called kinetic energy of the mode $E$, and
the dimensionless first-order solid rotation coefficient $C_{kl}$.
As is standard, our equilibrium stellar models are perfectly spherical,
and each mode is $2l+1$ fold-degenerate in eigenfrequency.

For asteroseismic studies, the most important quantities derived from
theoretical calculations are of course the periods of the modes. These
are sensitive to the global structural parameters of a model which
we seek to infer for a real star through a comparison with a set of
observed periods. Because nonadiabatic effects on the periods are
small (but included in our calculations), asteroseismology could very
well be carried out at the level of the adiabatic approximation only.
However, the stability coefficient, a purely nonadiabatic quantity,
also provides useful information on the mode driving mechanism. A
positive value of $\sigma_{I}$ indicates that a mode is damped (or
{}``stable''), and therefore that it should not be observable. A
negative value of $\sigma_{I}$ indicates that the mode is excited
(or {}``unstable'') in the model, and consequently that it may reach
observable amplitudes. As illustrated in Table \ref{cap:Best-fit-model},
modes are excited within a band of periods associated with low-order
acoustic modes. This is typical of all our models of rapid sdB pulsators.
In terms of asteroseismology, it is of course within that bandpass
that one would expect to find the observed periods of a real EC14026
star and this important stability information justifies the use of
full nonadiabatic calculations in our case. For its part, the kinetic
energy is a secondary quantity in the present context. It provides
a measure of the required energy to excite a mode of a given amplitude
at the surface of a star. Since this is a normalized quantity, only
its relative amplitude from mode to mode is of interest. The kinetic
energy is a useful diagnostic tool for determining where pulsation
modes are formed. Larger values of $E$ imply that such modes probe
deeper into the star, in higher density regions. The kinetic energy
bears the signature of mode trapping/confinement, which are caused
by partial reflexions of propagating waves onto thin chemical transition
zones (between the pure He-core and the H-rich envelope, mainly).
Finally, the rotation coefficient is useful for interpreting the fine
structure in the Fourier domain in terms of slow rotation that lifts
the $2\ell+1$ fold-degeneracy of the mode periods. At the frequency
resolution reached in our observations of \object{PG~1219+534}, no such structures
have been detected, however.

\subsection{The Quest for the Optimal Model of PG~1219+534}

The search for the optimal model solution (or solutions) that can
best reproduce the nine identified periods of \object{PG~1219+534} was initiated
with our multimodal genetic algorithm (GA) optimization code. As mentioned
briefly in the previous subsection, the search is carried out in the
four-dimensional space defined by the main parameters $T_{{\rm eff}}$,
$\log g$, $M_{*}$, and $\log q({\rm H})$ that specify the structure
of a sdB star with our second-generation models. The GA code, designed
for efficient explorations of a wide parameter space domain, was first
used to localize potential regions were solutions may exist. For that
purpose, initial boundaries of the search domain were defined
as follows : 37,000 K $\ge T_{{\rm eff}}\ge$ 28,000 K, $5.65\le\log g\le5.95$
for the surface parameters, and $-5.0\le\log q({\rm H})\le-2.0$,
$0.30\le M_{*}/M_{\odot}\le0.53$ for the structural parameters.

These limits were loosely set according to current spectroscopic estimates
for the atmospheric parameters $T_{{\rm eff}}$ and $\log g$ of \object{PG~1219+534}.
We emphasize the fact that any eventual determination of the atmospheric
parameters of \object{PG~1219+534} through asteroseismic means has
to be consistent with our spectroscopic estimates. Since the latter
are based on standard, well-understood techniques (see, however, the
discussions of \citealt{1997fbs..conf..433W} and \citealt{Heber2004} on possible systematic
errors introduced by these methods when applied to sdB stars), the
validity of our pulsating models would have to be questioned if the
results show otherwise. Constraints on the two other parameters $M_{*}$
and $\log q({\rm H})$ rely on stellar evolution theory. The evolutionary
calculations of \citet{1993ApJ...419..596D} indicate that hot B subdwarfs
are core helium burning stars on the Extreme Horizontal Branch that
evolve as ``AGB-Manqu\'e" objects. According to these calculations,
their possible masses are found in a narrow range $0.40-0.43\lesssim M_{*}/M_{\odot}\lesssim0.53$,
with a somewhat uncertain lower limit related to the minimum mass
required to ignite helium and a more sharply defined upper limit above
which the models evolve to the AGB. More recently, however, \citet{2003MNRAS.341..669H}
suggested, in their investigation of various binary evolution scenarios
for the formation of sdB stars, a somewhat larger distribution of
masses among these stars. Although these authors derived mass distributions
which are strongly peaked around the canonical value of $\sim0.47$
$M_{\odot}$, they suggest that the mass of an sdB star could be as
low as $\sim0.30$ $M_{\odot}$ if formed through the Roche lobe overflow
channel or as high as $\sim0.70$ $M_{\odot}$ if formed through the
merger channel. To explore these possibilities with asteroseismology,
we adopted a generous lower mass limit of $0.30$ $M_{\odot}$ but
kept $0.53$ $M_{\odot}$ as our upper boundary (results from test
calculations for models with higher masses will be briefly discussed,
though). Finally, the range of possible values for the last parameter
$\log q({\rm H)}$, which is related to the mass of the H-rich envelope,
was chosen according to the work of \citet{1993ApJ...419..596D} in
order to fully map the region of the $\log g-T_{{\rm eff}}$ plane
where sdB stars are found.

For the pulsation calculations step, we considered all the modes (be
it of $p$-, $f$-, or $g$-type) of degree $\ell=0$ up to $\ell=3$
with periods in the range $70-500$ s, i.e., covering amply the range of
periods observed in \object{PG~1219+534}. The usual argument to limit
the value of $\ell$ above which modes are no longer observable is
geometric cancellation on the visible disk of the star. A quantitative
expression of the visibility argument used to restrict the search
to low values of $\ell$ is obtained by evaluating numerically the
geometric factor given, for example, by equation (B7) of \citet{1995ApJS...96..545B}.
Using an Eddington limb darkening law in that formulation gives visibility
factors with values of 1.000, 0.7083, 0.3250, 0.0625, 0.0208, 0.0078,
0.0078, 0.0023, 0.0039, 0.0009, 0.0023, 0.0005, and 0.0015, respectively,
for $\ell=0$, 1, 2, 3, 4, 5, 6, 7, 8, 9, 10, 11, and 12. This cancellation
effect naturally favors \emph{a priori} values of $\ell=0$, 1, and
2 in pulsating stars in general. We note, however, that the mode densities
and period distributions seen in several well-observed EC14026 stars
such as \object{KPD 1930+2752}, \object{PG 1605+072}, \object{PG
  1047+003}, and \object{PG 0014+067}
force us to consider modes with higher $\ell$ values ($\ell=3$, or
even $\ell=4$ in some cases). Otherwise, there would be not enough
theoretical modes available in the observed period window to account
for the observations. To justify further the (small) heresy that
consists of considering modes of degree up to $\ell=3$ or 4 (although,
again, observational evidence already suggests that such modes are
effectively seen), we stress that our growing observational experience
with EC14026 pulsators indicates clearly that intrinsic mode amplitudes
in these stars are not the same from mode to mode. In addition, these
amplitudes may vary significantly on a monthly or yearly basis, some
modes being visible one season but disappearing
the next one. Therefore, one should be cautious with the simple
geometric cancellation argument which is based on the implicit assumption
that intrinsic amplitudes of all modes are the same. We suggest instead,
at least for EC14026 stars, that the limit in $\ell$ at which modes
cease to be observable due to cancellation effects cannot be a strict
threshold, since it also depends on the intrinsic amplitude of the
modes considered individually. For the purpose of asteroseismology,
we choose to consider modes from $\ell=0$ up to the minimum value
of the degree $\ell$ needed to account for the mode density
in the observed period range. For the \object{PG~1219+534} periods
identified in Table~\ref{cap:Identified-frequencies-for},
modes up to $\ell=3$ are needed. Of course, we cannot rule
out completely the presence of the odd $\ell=4$ modes (or perhaps
even a mode of higher degree) with an unusually high intrinsic amplitude
in the light curve of \object{PG~1219+534}. Our approach explicitly
excludes this possibility, however, and we seek to interpret the period
distribution in that star solely in terms of $\ell=0-3$ modes.

\begin{figure*}
\begin{center}\includegraphics[%
  scale=0.49]{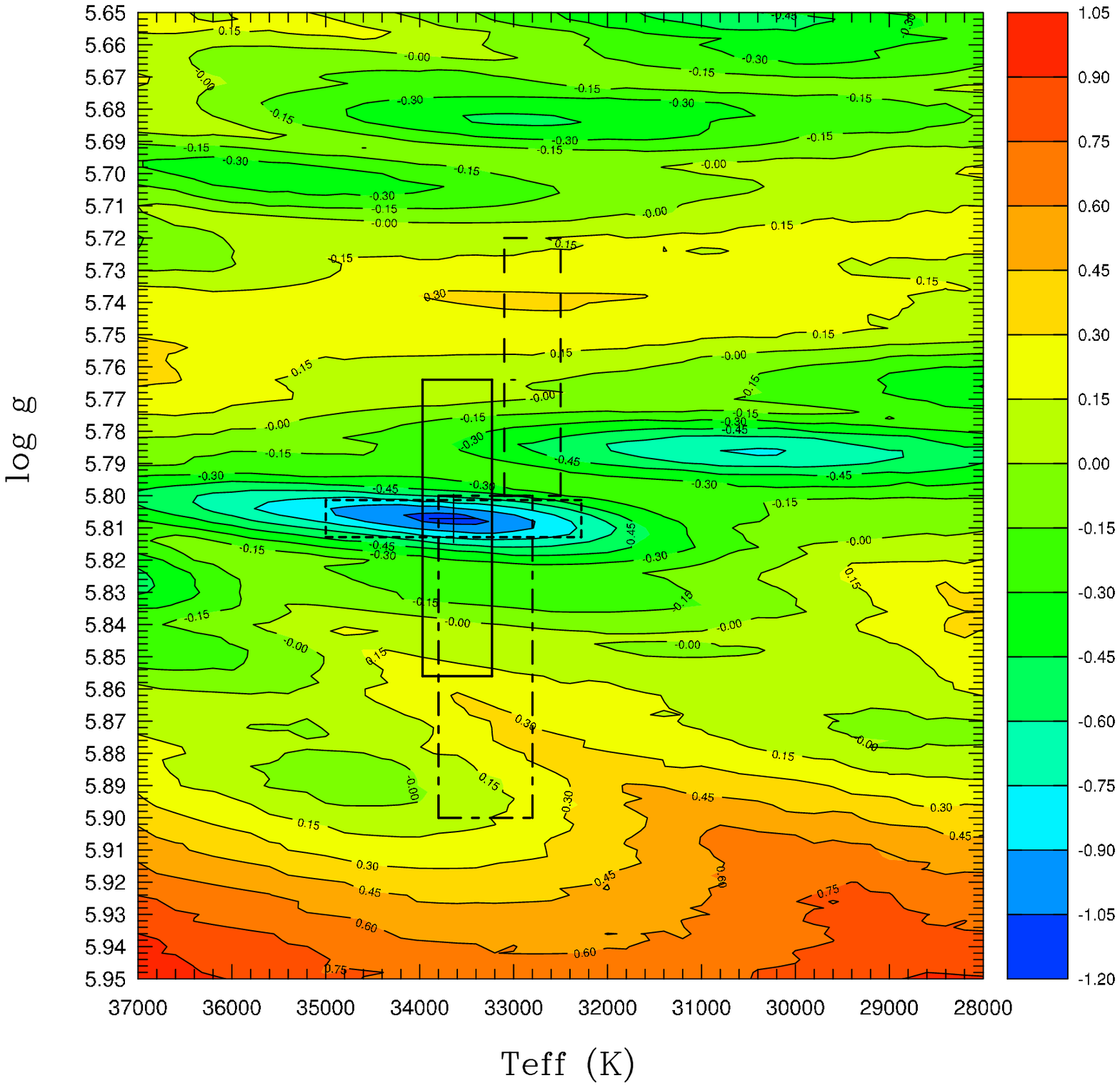}
  \hspace{5mm}
  \includegraphics[%
  scale=0.49]{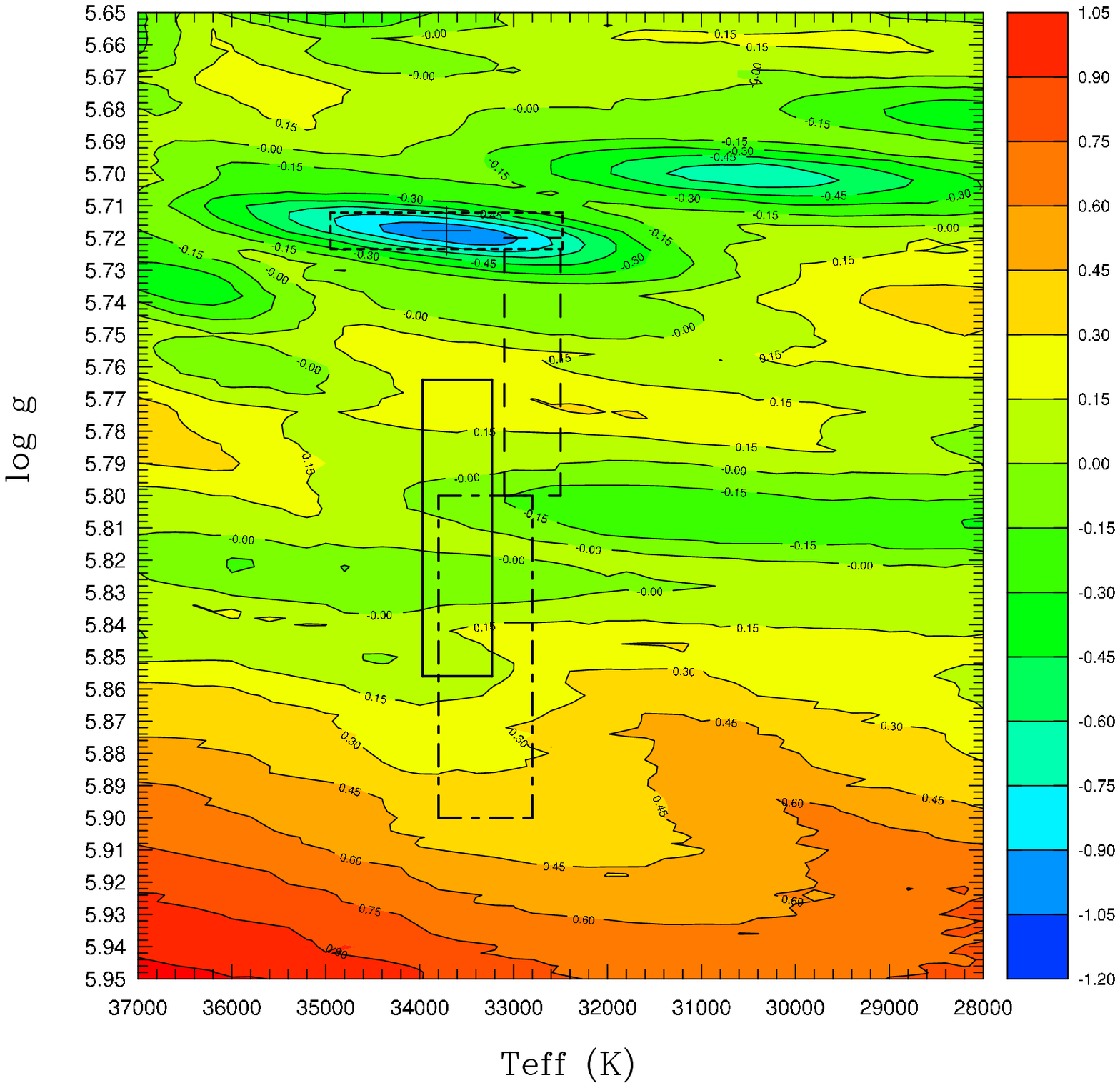}\end{center}

\caption{\emph{Left panel}: Slice of the $\overline{\chi}^{2}$-function (in
logarithmic units) along the $\log g-T_{{\rm eff}}$ plane at fixed
parameters $M_{*}$ and $\log q({\rm H})$ set to their optimal values
found for the first best-fit model solution ($M_{*}=0.4570$ $M_{\odot}$
and $\log q({\rm H})=-4.3448$). The solid-line rectangle shows
our spectroscopic estimate with its uncertainties for the atmospheric parameters
of \object{PG~1219+534}, while the dashed-line and dashed-dotted-line rectangles
represent the \citet{1999MNRAS.305...28K} and \citeauthor{2000A&A...363..198H}
(\citeyear{2000A&A...363..198H}; from NLTE H+He model atmospheres) spectroscopic measuremen
ts,
respectively.
The dotted-line rectangle shows the estimated
error box associated with our asteroseismic determination of the parameters
$\log g$ and $T_{{\rm eff}}$ (see text). The black cross indicates
the position of the best-fit solution. \emph{Right panel}: Same as
\emph{left panel} but for the second best-fit model solution (the
fixed parameters are set to the optimal values $M_{*}=0.4680$ $M_{\odot}$
and $\log q({\rm H})=-3.8142$). \label{cap:logg-teff}}
\end{figure*}

\begin{figure*}
\begin{center}\includegraphics[%
  scale=0.49]{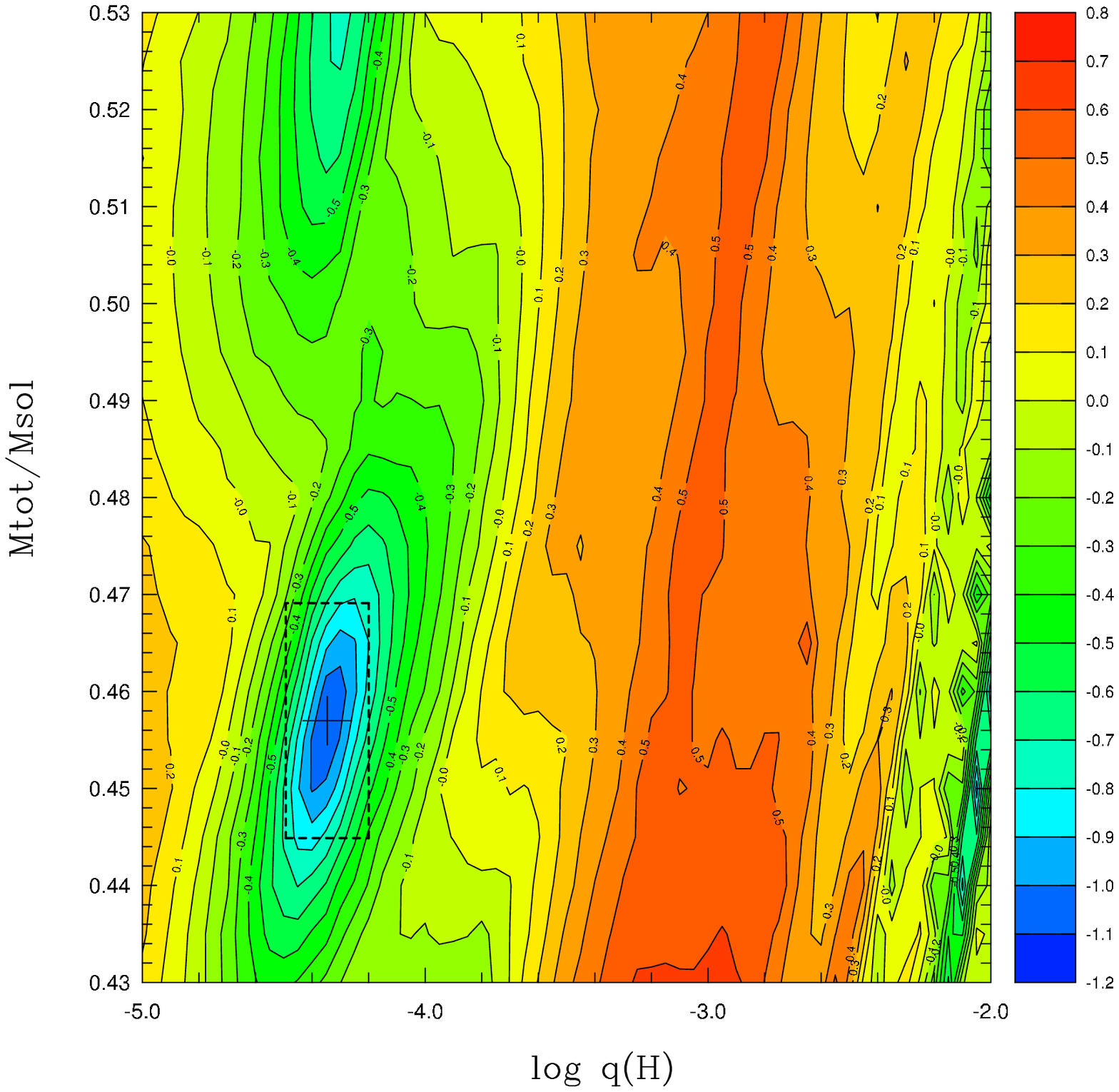}
  \hspace{5mm}
  \includegraphics[%
  scale=0.49]{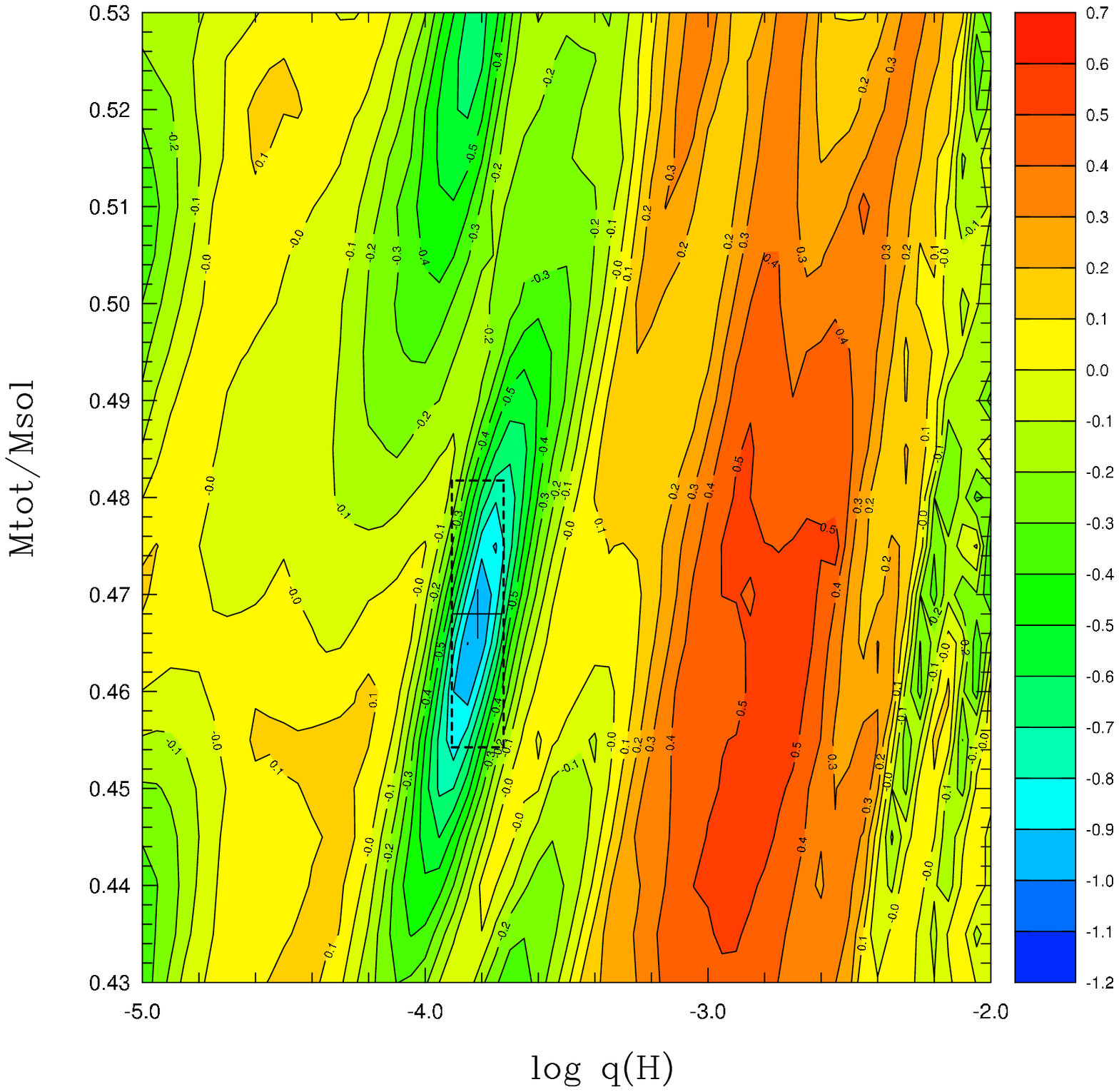}\end{center}

\caption{\emph{Left panel}: Slice of the $\overline{\chi}^{2}$-function (in
logarithmic units) along the $M_{*}-\log q({\rm H})$ plane at fixed parameters
$T_{{\rm eff}}$ and $\log g$ set to their optimal values found for
the first best-fit model solution ($T_{{\rm eff}}=$ 33,640 K and
$\log g=5.8071$). The dotted-line rectangle shows the estimated
error box associated with our asteroseismic determination of the parameters
$\log q({\rm H})$ and $M_{*}$ (see text). The black cross indicates
the position of the best-fit solution considered. \emph{Right panel}:
Same as \emph{left panel} but for the second best-fit model solution
(the fixed parameters are set to the optimal values $T_{{\rm eff}}=$
33,715 K and $\log g=5.7178$).\label{cap:lqh-mass}}
\end{figure*}

Within the search domain specified, the GA code identified two families
of solutions that best match the observed periods. A third family of potential
solutions was also spotted near the low-gravity edge of the search domain, but was
immediately rejected as being highly inconsistent with the available
spectroscopic measurements (see discussion below).
These model solutions are essentially equivalent in terms of quality of fit, indicating that
best-fitting the observed periods only leads to a degenerate situation (there is
another source of solution degeneracy which will be discussed below).
The optimal model corresponding to the first family of solutions was
found at $T_{{\rm eff}}=$ 33,640 K, $\log g=5.8071$, $\log q({\rm H})=-4.3448$,
and $M_{*}=0.4570$ $M_{\odot}$ with a $\overline{\chi}^{2}$ value
of 0.0784. The model associated with the second family of solutions
was localized at $T_{{\rm eff}}=$ 33,715 K, $\log g=5.7178$, $\log
q({\rm H})=-3.8142$,
and $M_{*}=0.4680$ $M_{\odot}$ with a $\overline{\chi}^{2}$ value
of 0.0878. Note that in the evaluation of $\overline{\chi}^{2}$,
we have chosen a global weight factor $\sigma_{i}=\sigma$, where
$\sigma$ is the inverse of the theoretical mode density, i.e., the
ratio of the width of the period window (here $500-70=430$ s) to
the number of theoretical modes in that window for a given model.
This choice was made in order to ease the automatic search of the
best-fit models in the parameter space by offsetting, at least in
part, the built-in bias in favor of models with a higher theoretical
mode density (generally, the models having a lower surface gravity).
Of course, this choice does not influence the positions of the minimum
values of $\overline{\chi}^{2}$ in the parameter space domain. Maps
displayed in Figure \ref{cap:logg-teff} and Figure \ref{cap:lqh-mass}
illustrate the complex shape of the $\overline{\chi}^{2}$-function
(shown as isocontours of constant value of $\log\overline{\chi}^{2}$)
in the vicinity of the two potential solutions, whose exact locations
according to the GA code are indicated by black crosses. These figures
respectively show slices of this function along the $\log g-T_{{\rm eff}}$
plane (at fixed parameters $M_{*}$ and $\log q({\rm H})$ set to
their optimal values) and along the $M_{*}-\log q({\rm H})$ plane
(at fixed parameters $T_{{\rm eff}}$ and $\log g$ set to their optimal
values). Best fitting models corresponding to low values of
$\overline{\chi}^{2}$ appear as dark blue regions, while red areas
indicate regions of the model parameter space with high values of
$\overline{\chi}^{2}$, i.e., where theoretical periods computed from
models do not fit well the observed periods. Considering the logarithmic
scale used to represent the merit function on these plots, we point out
that the blue regions correspond to well-defined minima.


If one family of solutions among the two identified cannot be preferred
over the other one simply on the basis of the $\overline{\chi}^{2}$
value alone, additional constraints provided by spectroscopic measurements
of the atmospheric parameters of \object{PG~1219+534} come in handy
for deciding which of these two best-fitting regions corresponds to
the {}``correct'' one. The values for $\log g$ and $T_{{\rm eff}}$
given by \citet{1999MNRAS.305...28K} and their associated errors
(shown as the dashed-line rectangle in Figure \ref{cap:logg-teff})
do not permit us to clearly discriminate between the two solutions.
However, the improved measurements obtained from our medium-resolution,
high signal-to-noise MMT spectrum (shown as the solid-line rectangle
in Figure \ref{cap:logg-teff}) allows us to unambiguously decide
that the solution at higher surface gravity (i.e., the first family
of solutions) is the preferred one. The spectroscopic values adopted
for \object{PG~1219+534} by \citet{2000A&A...363..198H} suggests
an even higher surface gravity (at $\log g=5.95\pm0.10$), thus further
dismissing the second family of solutions as a viable choice. We find,
however, that no model with such a high surface gravity can reproduce
the period spectrum observed in \object{PG~1219+534}. A closer look
at the period spectra computed for high-gravity models indicates that
this is simply due to the fact that some of the observed periods would
have to be associated with low-order gravity waves. However, the density
of $g$-modes is too low to be compatible with the mode density observed
in that star, and matching simultaneously several periods with $g$-modes
is impossible. Nonetheless, we again stress that the atmospheric values
derived by \citet{2000A&A...363..198H} from NLTE H+He model atmospheres
(shown as the dashed-dotted line rectangle\footnote{Note that
\citet{2000A&A...363..198H} do not provide error estimates specifically
associated with the parameters of \object{PG~1219+534} derived from their
NLTE H+He model atmospheres. We have therefore adopted $\Delta\Teff=500$ K
and $\Delta\log g=0.05$ as realistic representative values.}
in Figure \ref{cap:logg-teff}),
which are presumably the most realistic in terms of the physics involved,
are in much closer agreement with our own spectroscopic estimates and do not
conflict with asteroseismology.
In light of the various spectroscopic measurements,
we therefore adopt the first identified family of models, located
near $\log g\sim5.81$, as the best candidate solution that would
reproduce simultaneously the nine periods observed in \object{PG~1219+534}.
We point out at this stage that beyond the best-fit model degeneracy
discussed above, a second more subtle type of degeneracy of the asteroseismic
solution among the chosen family of models does exist, however. This
degeneracy occurs when a change in one of the model parameters is
almost exactly compensated by a change in another model parameter
such that the computed periods remain unmodified. This phenomenon
occurs in our analysis of \object{PG~1219+534} at two levels.

\begin{figure*}
\begin{center}\includegraphics[%
  scale=1.0]{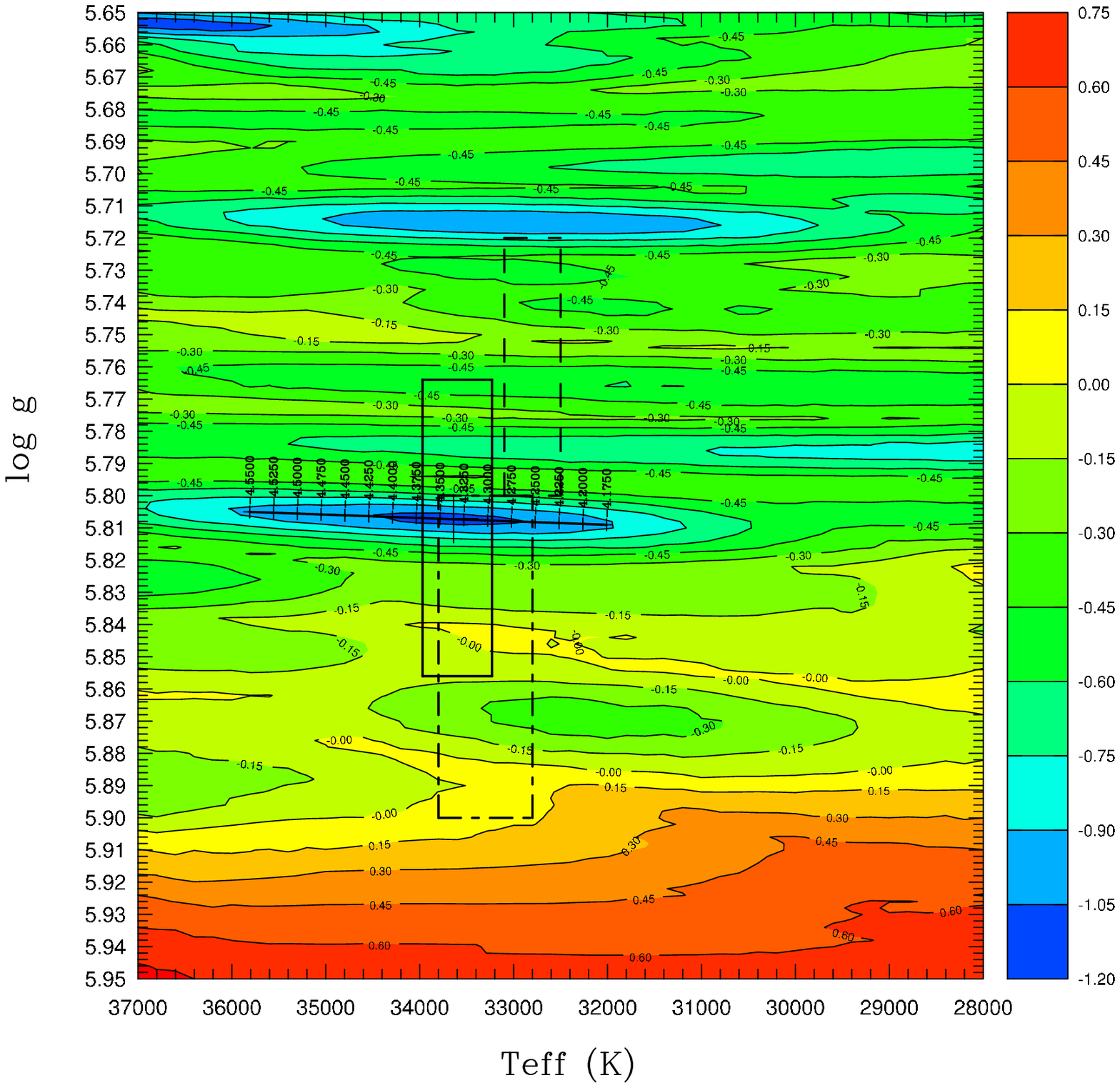}\\
\end{center}

\caption{Slice of the ``projected'' $\overline{\chi}^{2}$-function (in
logarithmic units) along the $\log g-T_{{\rm eff}}$ plane at fixed
parameter $M_{*}$
set to its optimal value found for the chosen best-fit model solution
($M_{*}=0.4570$ $M_{\odot}$). The projected, $\log q({\rm H})$
parameter was varied between $-2.0$ and $-5.0$ (by steps of 0.025).
The labelled axis positioned along the valley of minimum $\overline{\chi}^{2}$
indicates the exact location of the local minimum of $\overline{\chi}^{2}$
for the value of $\log q({\rm H})$ discussed in the text.
The solid-line rectangle shows
our spectroscopic estimate with its uncertainties for the atmospheric parameters
of \object{PG~1219+534}, while the dashed-line and dashed-dotted-line rectangles
represent the \citet{1999MNRAS.305...28K} and \citeauthor{2000A&A...363..198H}
(\citeyear{2000A&A...363..198H}; from NLTE H+He model atmospheres) spectroscopic
measurements, respectively.
\label{cap:lg-teff-lqh}}
\end{figure*}

First, there is a small correlation between the $\log q({\rm H})$
and $T_{{\rm eff}}$ parameters. Starting from the derived best-fitting model,
changing the value of $\log q({\rm H})$ while keeping $M_{*}$ constant
and set to its optimal value generates, in the $\log g-T_{{\rm eff}}$
plane, a shift of the position of the local minimum which mainly follows
the effective temperature axis (a tiny shift also occurs in $\log g$).
This trend is illustrated with the map provided in Figure
\ref{cap:lg-teff-lqh},
which shows the {}``projection'' of the $\log q({\rm H})$ axis
onto the surface gravity-effective temperature plane. More precisely,
the value associated with each grid point shown on this map is the minimum
value found among all the values of the $\overline{\chi}^{2}$-function
computed at fixed $T_{{\rm eff}}$, $\log g$ (set to the values associated
with the considered grid point), and $M_{*}$ (set to its optimal
value, that is $M_{*}=0.4570$ $M_{\odot}$) but with the parameter
$\log q({\rm H})$ varying within the limits of the specified search
domain, i.e., between $\log q({\rm H)=-5.0}$ and $-2.0$. The various
families of solutions identified previously now appear simultaneously
as (blue) parallel valleys of low $\overline{\chi}^{2}$ in the $\log g$-$T_{{\rm eff}}$
plane. These correspond, however, to a different range of values for
the parameter $\log q({\rm H})$. The small labelled axis positioned
along the valley associated with the preferred solution indicates
the position of the local minimum of $\overline{\chi}^{2}$ as a function
of $\log q({\rm H})$. There is a clear, monotonic trend showing that
this minimum shifts from higher effective temperatures (e.g., $T_{{\rm eff}}\sim$
35,800 K for $\log q({\rm H})\sim-4.55$) to lower $T_{{\rm eff}}$
($\sim$ 32,000 K for $\log q({\rm H)\sim-4.175}$) as the envelope
mass of the star increases (i.e., $\log q({\rm H})$ increases). However,
at the same time, we note a degradation of the quality of the period
fit (i.e., an increase of the absolute value of $\overline{\chi}^{2}$
associated with the minimum) for values of $\log q({\rm H})$ that
departs from the optimal value uncovered. This indicates that the
degeneracy associated with the parameter $\log q({\rm H})$ has boundaries,
i.e, the optimal solution still occupies the center of a well defined
region of the parameter space. Interestingly, we note at this point
that best-fit model candidates for the observed periods of \object{PG~1219+534}
do all correspond to models with thin H-rich envelopes. In our search,
no acceptable asteroseismological fit has been found for models with
thick envelopes, i.e., with high values of the parameter $\log q({\rm H})$.

\begin{figure*}
\begin{center}\includegraphics[%
  scale=1.0]{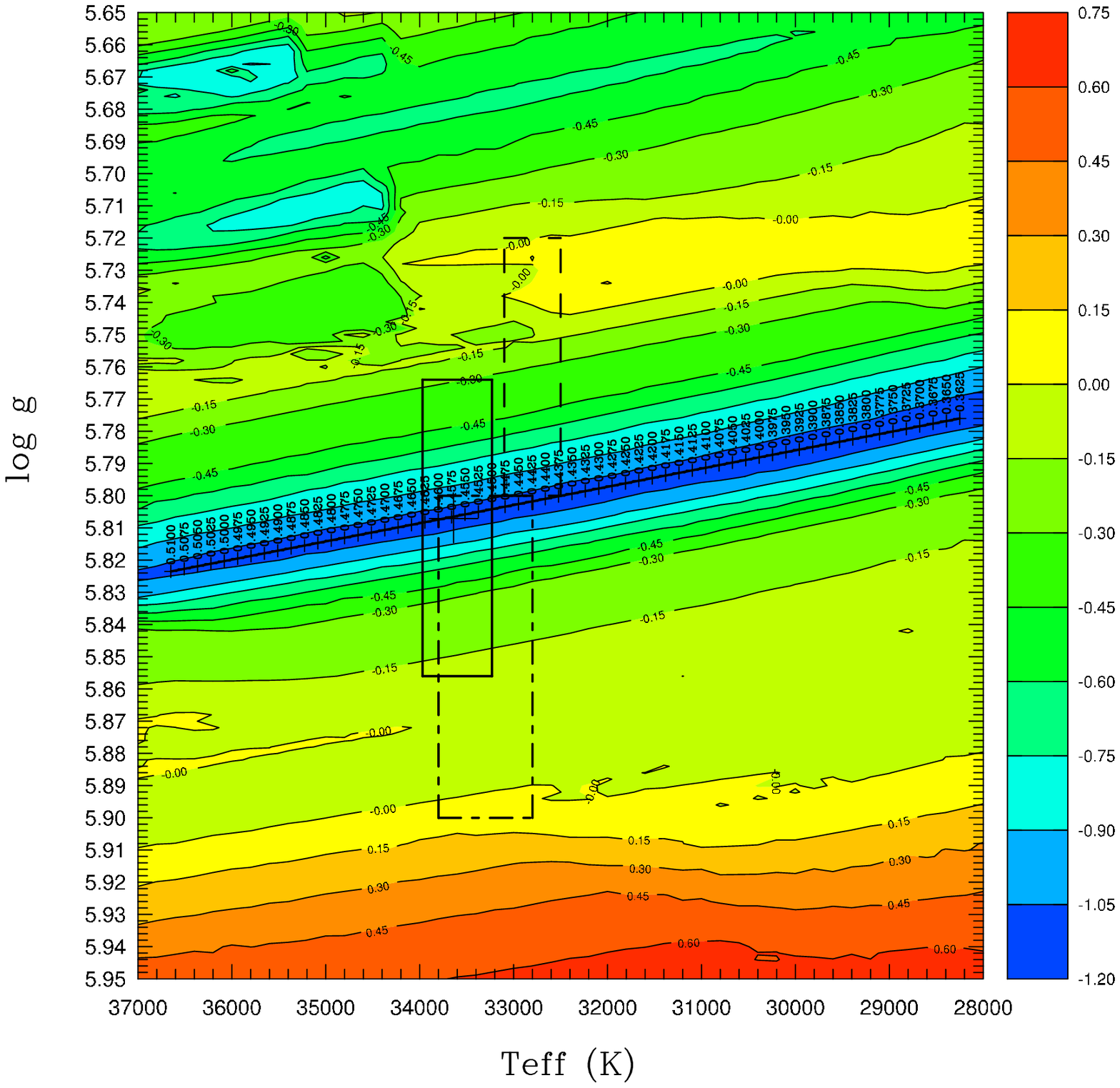}\\
\end{center}

\caption{Slice of the ``projected'' $\overline{\chi}^{2}$-function (in
logarithmic units) along the $\log g-T_{{\rm eff}}$ plane at fixed
parameter $\log q({\rm H})$
set to its optimal value found for the chosen best-fit model solution
($\log q({\rm H})=-4.3448$). The projected, $M_{*}$ parameter was
varied between $0.30$ $M_{\odot}$ and $0.53$ $M_{\odot}$ (by steps
of 0.0025 $M_{\odot}$). The labelled axis positioned along the valley
of minimum $\overline{\chi}^{2}$ indicates the exact location of
the local minimum of $\overline{\chi}^{2}$ for the value
of $M_{*}$ discussed in the text.
The solid-line rectangle shows
our spectroscopic estimate with its uncertainties for the atmospheric parameters
of \object{PG~1219+534}, while the dashed-line and dashed-dotted-line rectangles
represent the \citet{1999MNRAS.305...28K} and \citeauthor{2000A&A...363..198H}
(\citeyear{2000A&A...363..198H}; from NLTE H+He model atmospheres) spectroscopic
measurements, respectively.
\label{cap:lg-teff-mtot}}
\end{figure*}

A similar map was constructed to visualize the {}``projection''
of the $M_{*}$-axis onto the $\log g-T_{{\rm eff}}$ plane. It is
shown in Figure~\ref{cap:lg-teff-mtot}. The parameter $\log q({\rm H})$
was kept constant (set to its optimal value of $\log q({\rm H})=-4.3448$)
and the total mass was varied within the limits we have defined for
the search domain, i.e., between $M_{*}=0.53$ $M_{\odot}$ and $M_{*}=0.30$
$M_{\odot}$. The map indicates clearly that a correlation also exists
between the parameter $M_{*}$ and the parameters $T_{{\rm eff}}$
and, to a lesser extent, $\log g$. A change in $M_{*}$ generates
a shift in both $T_{{\rm eff}}$ and $\log g$ of the position of
the $\overline{\chi}^{2}$ minimum. However, contrary to the case
of the $\log q({\rm H})$ parameter discussed previously, we observe
no degradation of the quality of the period fit over the entire range
considered for the $M_{*}$ parameter. Clearly in the case of \object{PG~1219+534},
the impact of changing the value of $M_{*}$ on the theoretical modes
fitted to the observed periods can be completely compensated by an
adequate change of both $T_{{\rm eff}}$ and $\log g$. This leads
to a line-degeneracy in the $\overline{\chi}^{2}$-function along
which models provide comparable best-fit solutions of the periods.
This is well illustrated in Figure \ref{cap:lg-teff-mtot} by the
presence of a long and flat valley of minimum value of $\overline{\chi}^{2}$
with no high-$T_{{\rm eff}}$ and low-$T_{{\rm eff}}$ boundaries.
We also mention that a similar line-degeneracy associated with the
second family of solutions, near $\log g\sim5.71$, does exist, although
it is not apparent in Figure \ref{cap:lg-teff-mtot} as it requires
a larger optimal value for the $\log q({\rm H})$ parameter. The labelled
axis plotted along the valley of minimum $\overline{\chi}^{2}$ indicates
the position of the best-fit solution in the $\log g-T_{{\rm eff}}$
plane for the given value of the total mass. The correlation between
$M_{*}$ and ($T_{{\rm eff}}$, $\log g$) is monotonic, as the effective
temperature and surface gravity of the solution decreases when the
total mass decreases. For instance, a value of $M_{*}=0.5075$ $M_{\odot}$
places the minimum of $\overline{\chi}^{2}$ near $T_{{\rm eff}}\sim$
36,600 K and $\log g\sim5.822$, while a value of $M_{*}=0.3650$
$M_{\odot}$ places it near $T_{{\rm eff}}\sim$ 28,400 K and $\log g\sim5.796$.
Note that the correlation with the effective temperature is much stronger
than the correlation with the surface gravity parameter. Of course,
this trend extends beyond the limits provided in Figure \ref{cap:lg-teff-mtot},
and models with lower (higher) masses lead to even cooler (hotter)
solutions. Again, one cannot discriminate between the models on the
basis of their $\overline{\chi}^{2}$-value alone, and spectroscopy
must be used to lift the degeneracy. The effective temperature derived
from the analysis of our medium-resolution, high signal-to-noise MMT
spectrum (the solid-line rectangle in Figure \ref{cap:lg-teff-mtot})
allows us to select the appropriate section along the line of degeneracy
which corresponds to the {}``correct'' solution. Of course, the
necessity to rely on the spectroscopic measurement of $T_{{\rm eff}}$
to uniquely derive the total mass of \object{PG~1219+534} from
asteroseismological means indicates, unfortunately, that $M_{*}$ cannot
be measured independently
of $T_{{\rm eff}}$ for that star. Since, the optimal model first
isolated by our GA code with parameters $T_{{\rm eff}}=$ 33,640
K, $\log g=5.8071$, $\log q({\rm H})=-4.3448$, and $M_{*}=0.4570$
$M_{\odot}$ is consistent with the spectroscopic determination of
the effective temperature of \object{PG~1219+534}, we therefore adopt
it as our preferred solution that best matches the period spectrum observed
in this star.

\begin{table*}

\caption{Pulsation Properties of the Best-Fit Model Solution and Mode Identification.
\label{cap:Best-fit-model}}

\begin{center}\begin{tabular}{cccccccccl}
\hline
&
&
$P_{{\rm obs}}$&
$P_{{\rm th}}$&
$\sigma_{I}$&
Stability&
$\log E$&
$C_{kl}$&
$\Delta P/P$&
Comments\tabularnewline
$\ell$&
$k$&
(s)&
(s)&
(rad/s)&
&
(erg)&
&
(\%)&
\tabularnewline
\hline
0&
6&
...&
76.154&
$+5.530\times10^{-6}$&
stable&
39.984&
...&
...&
\tabularnewline
0&
5&
...&
85.013&
$-5.270\times10^{-5}$&
unstable&
40.028&
...&
...&
\tabularnewline
0&
4&
...&
98.416&
$-4.276\times10^{-5}$&
unstable&
40.389&
...&
...&
\tabularnewline
0&
3&
...&
107.574&
$-1.974\times10^{-5}$&
unstable&
40.743&
...&
...&
\tabularnewline
0&
2&
129.099&
129.406&
$-8.392\times10^{-6}$&
unstable&
41.003&
...&
$-0.24$&
$f_{6}$\tabularnewline
0&
1&
148.775&
148.579&
$-4.386\times10^{-7}$&
unstable&
42.055&
...&
$+0.13$&
$f_{4}$\tabularnewline
0&
0&
...&
165.145&
$-2.578\times10^{-7}$&
unstable&
42.063&
...&
...&
\tabularnewline
&
&
&
&
&
&
&
&
&
\tabularnewline
1&
7&
...&
74.783&
$+1.932\times10^{-5}$&
stable&
39.894&
0.0071&
...&
\tabularnewline
1&
6&
...&
84.243&
$-5.145\times10^{-5}$&
unstable&
39.999&
0.0067&
...&
\tabularnewline
1&
5&
...&
96.410&
$-3.634\times10^{-5}$&
unstable&
40.434&
0.0125&
...&
\tabularnewline
1&
4&
...&
105.468&
$-2.941\times10^{-5}$&
unstable&
40.572&
0.0120&
...&
\tabularnewline
1&
3&
128.077&
128.176&
$-8.586\times10^{-6}$&
unstable&
40.999&
0.0141&
$-0.08$&
$f_{1}$\tabularnewline
1&
2&
143.649&
143.327&
$-8.108\times10^{-7}$&
unstable&
41.846&
0.0280&
$+0.22$&
$f_{2}$\tabularnewline
1&
1&
...&
164.511&
$-2.909\times10^{-7}$&
unstable&
42.018&
0.0176&
...&
\tabularnewline
&
&
&
&
&
&
&
&
&
\tabularnewline
2&
7&
...&
73.101&
$+4.804\times10^{-5}$&
stable&
39.746&
0.0083&
...&
\tabularnewline
2&
6&
{[}82.261{]}&
83.016&
$-4.504\times10^{-5}$&
unstable&
39.981&
0.0095&
$[-0.92]$&
$[f_{10}]$\tabularnewline
2&
5&
...&
92.659&
$-3.111\times10^{-5}$&
unstable&
40.445&
0.0223&
...&
\tabularnewline
2&
4&
...&
103.257&
$-4.166\times10^{-5}$&
unstable&
40.420&
0.0145&
...&
\tabularnewline
2&
3&
122.408&
124.141&
$-7.354\times10^{-6}$&
unstable&
41.081&
0.0413&
$-1.42$&
$f_{5}$\tabularnewline
2&
2&
135.160&
135.309&
$-3.273\times10^{-6}$&
unstable&
41.331&
0.0462&
$-0.11$&
$f_{8}$\tabularnewline
2&
1&
...&
163.248&
$-3.474\times10^{-7}$&
unstable&
41.953&
0.0237&
...&
\tabularnewline
2&
0&
...&
196.428&
$-3.626\times10^{-11}$&
unstable&
44.808&
0.4159&
...&
\tabularnewline
&
&
&
&
&
&
&
&
&
\tabularnewline
3&
7&
...&
71.517&
$+9.095\times10^{-5}$&
stable&
39.612&
0.0713&
...&
\tabularnewline
3&
6&
...&
80.835&
$-2.714\times10^{-5}$&
unstable&
40.013&
0.0218&
...&
\tabularnewline
3&
5&
...&
88.223&
$-3.861\times10^{-5}$&
unstable&
40.262&
0.0310&
...&
\tabularnewline
3&
4&
...&
101.203&
$-4.846\times10^{-5}$&
unstable&
40.348&
0.0209&
...&
\tabularnewline
3&
3&
...&
115.569&
$-7.059\times10^{-6}$&
unstable&
41.145&
0.0822&
...&
\tabularnewline
3&
2&
133.516&
130.982&
$-6.919\times10^{-6}$&
unstable&
41.059&
0.0317&
$+1.90$&
$f_{3}$\tabularnewline
3&
1&
158.789&
160.146&
$-4.134\times10^{-7}$&
unstable&
41.910&
0.0678&
$-0.85$&
$f_{9}$\tabularnewline
3&
0&
172.214&
171.664&
$-2.565\times10^{-8}$&
unstable&
42.888&
0.1774&
$+0.32$&
$f_{7}$\tabularnewline
\hline
\end{tabular}\end{center}
\end{table*}

Of interest in the context of binary evolution scenarios to form hot
B subdwarf stars, we find that the low-mass (high-mass) models consistent
with the observed periods of \object{PG~1219+534} would require effective
temperatures which are significantly cooler (hotter) than currently
measured by spectroscopic means. Therefore, those solutions can clearly
be rejected on the basis of consistency between spectroscopy and
asteroseismology. Remarkably, the mass derived for \object{PG~1219+534},
which is compatible with the spectroscopically measured effective
temperature, is found to be consistent with the canonical value for
Extreme Horizontal Branch objects.

\subsection{Period Fit and Mode Identification}

\begin{figure}
\begin{center}\includegraphics[%
  bb=50bp 144bp 592bp 718bp,
  clip,
  scale=0.45]{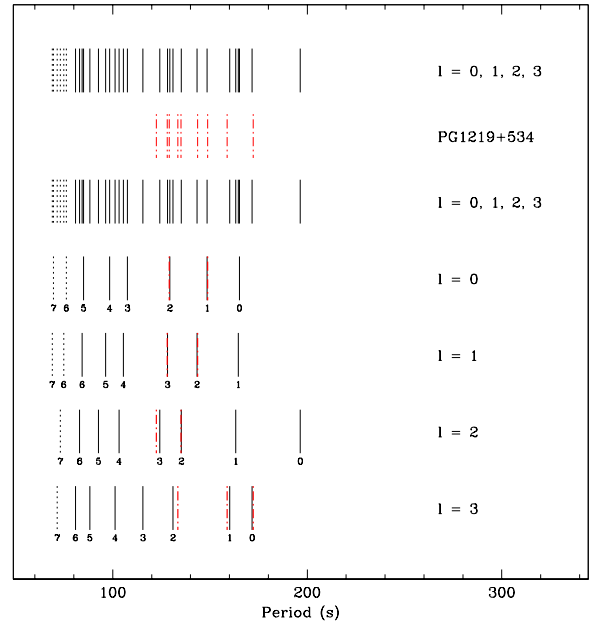}\end{center}

\caption{Comparison of the observed period spectrum of \object{PG~1219+534}
(\emph{thick dashed-dotted-line red segments}) with the theoretical
pulsation spectrum of the optimal model. For the latter spectrum,
\emph{solid-line segments} indicate excited modes, while \emph{dotted-line
segments} correspond to damped modes. All pulsation modes with $\ell=0$,
1, 2, and 3 in the period interval $70-200$ s are illustrated. The
values of the radial order index $k$ are also indicated for each
mode. These modes are all acoustic waves (including the $f$-modes).
The $g$-modes have periods that fall outside the range of interest
for \object{PG~1219+534}. \label{cap:Mode-identification-for}}
\end{figure}

The optimal model finally isolated in the previous subsection provides
the best simultaneous period match of the nine periods detected in
\object{PG~1219+534} and leads to the identification of the pulsation
modes involved in the luminosity variations observed in that star.
We recall that our global optimization method applied to the asteroseismic
analysis of \object{PG~1219+534} does not rely on any previous assumption
concerning the values of the degree $\ell$ and the radial order $k$
of the modes being observed. The mode identification appears instead
naturally -- and somewhat objectively -- as a solution (or prediction)
of the double-optimization procedure. Details on the derived mode
identification and period fit for \object{PG~1219+534} are given
in Table \ref{cap:Best-fit-model} (a graphical representation of
it is also shown in Figure \ref{cap:Mode-identification-for}). In
addition to the quantities that reflect the properties of the nonradial
modes computed from the best-fit model (previously described in Subsection
4.2), Table \ref{cap:Best-fit-model} provides the derived distribution
of the observed periods ($P_{{\rm obs}}$) as these were matched to
the theoretical periods ($P_{{\rm th}}$). Again, this distribution
is the one that minimizes $\chi^{2}$, the sum of the squared difference
between the observed periods and their assigned theoretical periods
for that model. The relative difference in period $\Delta P/P$ (in
\%) for each pair ($P_{{\rm obs}},P_{{\rm th}}$) is also given in
this table.

We find that the simultaneous fit of the nine periods of \object{PG~1219+534}
obtained with the optimal model retained is excellent by current
standards. The average relative dispersion between
the fitted periods is $\overline{\Delta P/P}\sim0.6\%$
and the worst difference is less than $\sim2\%$. On an absolute scale,
the average dispersion between the periods is less than $\sim0.8$
s. This accurate and simultaneous fit of nine periods constitutes, by
itself, a remarkable result in the field of asteroseismology. It is
comparable in quality (and even slightly better due, in part, to the
improved tools used to isolate the optimal model) to the results achieved
by Brassard et al. (2001) for the star \object{PG 0014+067}.
Yet, we emphasize the fact that the accuracy at which the observed
periods are measured is approximately one order of magnitude better
than the mean period dispersion achieved by our optimal fit. This
suggests that, as good as this fit appears to be by current standards
of asteroseismology, our equilibrium models describing the structure
of sdB stars still suffer from imperfections that leave significant
room for improvement, on an absolute scale, in reproducing the current
observations. This is, of course, one of the goals to pursue in future
asteroseismic studies of sdB pulsators. Interestingly, the marginal
detection of an independent period at $82.261$ s ($f_{10}$, shown
within brackets in Table \ref{cap:Best-fit-model}), although it was
not used at all to constrain the search for the optimal model, can
indeed, after the fact, be associated with a theoretical mode without
degrading the overall quality of the period fit. This result indicates
that this period may actually correspond to a real mode excited at
very low amplitude in \object{PG~1219+534}. In addition, we find
that the 9 observed periods (10, if we consider $f_{10}$ as real)
all fall within the predicted band of instability, as it is clearly
apparent in Table \ref{cap:Best-fit-model}. Hence, the model solution
derived from our global optimization procedure is also remarkably
consistent with the prediction of nonadiabatic pulsation theory applied
to our second-generation models of pulsating hot B subdwarfs. The
observed periods are identified with radial ($\ell=0$) and nonradial
($\ell=1,$ 2, and 3) $f$- and $p$-modes with radial orders between
$k=0-3$ (note that the 82.261 s period, if real, would correspond
to a mode with $k=6$, i.e., at the short-period end of the band of
instability). Interestingly, the mode identification inferred from
the optimal model presented in Table \ref{cap:Best-fit-model} remains
unchanged along the line of degeneracy identified in the previous
subsection. Thus, the models found along this line of degeneracy can
indeed be considered as members of a family of solutions in this respect.
Also of interest, we find that the mode identification associated
with the second (rejected) family of solutions found near $\log g\sim5.71$
is similar to the identification proposed, except that all modes are
shifted by $\Delta k=+1$. This suggests a periodic behavior, with
a third family of acceptable best-fit solutions with modes now shifted by $\Delta k=+2$.
However, such a solution would correspond to a model with a surface gravity
lower than our bound of $\log g=5.65$ in our search volume in the parameter space,
a value in clear conflict with available spectroscopic estimates.

Obviously, from Table \ref{cap:Best-fit-model} and Figure \ref{cap:Mode-identification-for},
modes of degree up to $\ell=3$ are required in order to account for
the mode density observed in \object{PG~1219+534}. Moreover, all
the observed modes tend to cluster near the low radial order (low-$k$)
boundary of the predicted band of instability. Similar mode distributions
are actually seen in \object{PG 0014+067} (see \citealt{2001ApJ...563.1013B})
and in a few other well studied EC14026 pulsators. Often, however,
a few periods may be found which correspond to modes positioned near
the high-$k$ boundary of the instability band, thus leaving holes
of predicted but unseen modes in the period spectrum. A suggestion
is that, within the band of excited periods, energy powering up the
pulsations may be preferentially distributed among the modes of low
radial order. These modes would then reach observable amplitudes more
easily, and therefore would be preferentially seen in the light curves
of EC14026 pulsators. We also note that we find no strongly established
hierarchy of the mode amplitude as a function of $\ell$ for the nine
modes identified in Table \ref{cap:Best-fit-model}. One would naively
expect, at least statistically, some trend of amplitude attenuation
when $\ell$ increases, due to geometric cancellation effects. In
this context, we note that three of the four dominant periods are
indeed associated with modes of degree $\ell=0$ and 1, while periods
of lower amplitudes seem to be preferentially associated with modes
of degree $\ell=2$ and 3 (with the noticeable exception of the mode
$f_{3}$ identified as a $\ell=3$ mode). All these issues related
to mode amplitudes are, however, beyond the realm of linear theory
and only a nonlinear approach to the pulsation phenomenon will be
able to address properly these questions. In the meantime, a purely
observational approach, such as searching for the ``missing'' modes at
still higher sensitivity than we could reach with the CFHT, for example
using one nigth of \noun{Gemini} time, could go a long way toward solving this
puzzle.

Finally, we stress that the $(k,\ell)$ values attributed to the periods
observed in \object{PG~1219+534} constitute \emph{a prediction of
our seismic analysis which is amenable to independent observational
tests}. A promising avenue for independent mode identification resides
in multicolor photometry (as opposed to one color, ``white light''
photometry used, for instance, in this study). It is well known from
stellar pulsation theory that the apparent amplitude of a nonradial
oscillation mode is a function of wavelength. Moreover, how this amplitude
changes with wavelength depends on the degree $\ell$ of the mode.
Consequently, measuring relative amplitude ratios of modes observed
at different wavelength allows, in principle, the identification of
their degree $\ell$. Such independent evaluations of the geometry
of the modes would obviously provide important tests of our seismic
analyses. Some very encouraging results based on that technique have been
presented recently by Jeffery et al. (2004), including the suggested
identification of a mode with $\ell=4$ in the hot pulsating sdB star
KPD 2109+4401.

\subsection{Structural Parameters of \object{PG~1219+534}}

The model that has been isolated and which best represents the observed
properties of \object{PG~1219+534}, from both the spectroscopy and
asteroseismology standpoints, leads to the determination of the fundamental
parameters that define the structure of this hot pulsating B subdwarf.
A first set of primary quantities corresponding to the main model
parameters is naturally derived from this exercise. These primary
parameters are the effective temperature $T_{{\rm eff}}$, the logarithm
of the surface gravity $\log g$, the total mass $M_{*}$, and the
mass of the hydrogen-rich envelope through the quantity $\log q({\rm H})\simeq\log(M_{{\rm env}}/M_{*})$.
A set of secondary parameters then follows from the values obtained
for the primary quantities. These secondary parameters are the radius
$R$ (as a function of $M_{*}$ and $g$), the luminosity $L$ (as
a function of $T_{{\rm eff}}$ and $R$), the absolute magnitude $M_{V}$
(as a function of $g$, $T_{{\rm eff}}$, and $M_{*}$ in conjunction with the use
of detailed model atmospheres), the distance from Earth $d$ (as a
function of $V$ and $M_{V}$). Limits on the rotation period $P_{{\rm rot}}$,
equatorial rotation velocity $V_{{\rm eq}}$(as a function of $R$
and $P_{{\rm rot}}$), and projected equatorial rotation velocity
$V_{{\rm eq}}\sin i$ could also be set due to the absence of fine
structure in the frequency spectrum of \object{PG~1219+534}, as multiplets
may just be unresolved with the time baseline provided by our observations.
The values and limits derived for all these quantities are summarized
in Table~\ref{cap:Inferred-Properties-of}.

\begin{table}

\caption{Inferred Properties of \object{PG~1219+534} ($V=13.24\pm0.20$).\label{cap:Inferred-Properties-of}}

\begin{center}{\scriptsize }\begin{tabular}{lcc}
\hline
{\scriptsize Quantity}&
{\scriptsize Asteroseismology}&
{\scriptsize Spectroscopy}\tabularnewline
\hline
\hline
{\scriptsize $\log g$}&
{\scriptsize $5.8071\pm0.0057$ (0.10\%)}&
{\scriptsize $5.810\pm0.046$ (0.79\%)}\tabularnewline
{\scriptsize $\Teff$ (K)}&
{\scriptsize 33,640 $\pm$ 1,360 (4.04\%)}&
{\scriptsize 33,600 $\pm$ 370 (1.10 \%)}\tabularnewline
{\scriptsize $M_{*}/M_{\odot}$}&
{\scriptsize $0.457\pm0.012$ (2.63\%)}&
{\scriptsize ...}\tabularnewline
{\scriptsize $\log(M_{{\rm env}}/M_{*})$}&
{\scriptsize $-4.254\pm0.147$ (3.46\%)}&
{\scriptsize ...}\tabularnewline
&
&
\tabularnewline
{\scriptsize $R/R_{\odot}$($M_{*}$, $g$)}&
{\scriptsize $0.1397\pm0.0028$ (2.00\%)}&
{\scriptsize ...}\tabularnewline
{\scriptsize $L/L_{\odot}$($\Teff$, $R$)}&
{\scriptsize $22.12\pm4.46$ (20.2\%)}&
{\scriptsize $22.01\pm1.85$ (8.41\%)}\tabularnewline
{\scriptsize $M_{V}(\textrm{$g$, $\Teff$, $M_{*}$)}$}&
{\scriptsize $4.61\pm0.12$ (2.60\%)}&
{\scriptsize $4.62\pm0.06$ (1.30\%)}\tabularnewline
{\scriptsize $d$($V$, $M_{V}$) (pc)}&
{\scriptsize $532\pm37$ (6.95\%)}&
{\scriptsize $531\pm23$ (4.33\%)}\tabularnewline
{\scriptsize $P_{{\rm rot}}$ (day)}&
{\scriptsize $\gtrsim3.4$}&
{\scriptsize ...}\tabularnewline
{\scriptsize $V_{{\rm eq}}$($R$, $P_{{\rm rot}}$) (km.s$^{-1}$)}&
{\scriptsize $\lesssim2.1$}&
{\scriptsize ...}\tabularnewline
{\scriptsize $V_{{\rm eq}}\sin i$ (km.s$^{-1}$)}&
{\scriptsize $\lesssim2.1$}&
{\scriptsize $\lesssim10$}\tabularnewline
\hline
\end{tabular}\end{center}
\end{table}

An essential ingredient of our asteroseismic study is an evaluation
of the uncertainties associated with the parameters derived for the optimal
model. The determination of these uncertainties is closely
linked to the behavior of the $\overline{\chi}^{2}$ hypersurface
in the vicinity of the best-fit solution (see Figures \ref{cap:logg-teff}
and \ref{cap:lqh-mass}). Following \citet{2001ApJ...563.1013B},
we can model the 4-dimensional $\overline{\chi}^{2}$-function in
terms of a quadratic expansion considering that the first-order derivatives
vanishe at the minimum and that the cross-terms in the second-order
derivatives also vanish, in a first approximation, if we consider
the four parameters involved to be independent. Clearly, that latter
statement is not exact, especially in light of the nature of the degeneracies
that have been uncovered in our search for the best-fit solution.
These complications will be further discussed below, but we will show
that they do not, however, significantly affect the evaluation of
the uncertainties associated with the parameters. This simplified
approach leads to the following relationship between the variation
$\Delta\overline{\chi}^{2}$ and the uncertainties, $\Delta a_{i}$,
on the parameters $a_{i}$ (where $a_{1}=\log g$, $a_{2}=T_{{\rm eff}}$,
$a_{3}=M_{*}$, and $a_{4}=\log q[{\rm H}]$)\begin{equation}
\Delta\overline{\chi}^{2}=\sum_{i=1}^{4}\frac{1}{2}\frac{\partial^{2}\overline{\chi}^{2}}{\partial a_{i}^{2}}\bigg|_{{\rm min}}(\Delta a_{i})^{2}\,\,\,\,.\label{eq:confidence-limit}\end{equation}
The second-order derivatives are evaluated at the position of the
optimal model and are estimated on the basis of the $\overline{\chi}^{2}$-values
found in the grids specifically constructed around this optimal model
(again, shown in Figures \ref{cap:logg-teff} and \ref{cap:lqh-mass}).
In order to obtain a conservative estimate of the uncertainty on a
given parameter, we assume that the variation $\Delta\overline{\chi}^{2}$
is due exclusively to the variation of that one parameter, thus leading
to the relationship\begin{equation}
\Delta a_{i}=\sqrt{2\Delta\overline{\chi}^{2}\bigg/\frac{\partial^{2}\overline{\chi}^{2}}{\partial a_{i}^{2}}\bigg|_{{\rm min}}}\,\,\,\,.\label{eq:errors}\end{equation}
 The problem now resides in evaluating the quantity $\Delta\overline{\chi}^{2}$.
The merit function $\overline{\chi}^{2}$ that we have defined to
measure the quality of the period fit is not a standard $\chi_{{\rm std}}^{2}$
for which usual formulae can apply to estimate the normal errors.
A renormalization of this function is needed and can be achieved following
the prescription of Press et al. (\citeyear{1985nreb.book.....V};
see also \citealt{1969nreb.book.....V}) assuming that we have obtained
a perfect fit of the observations. In that case, the standard $\chi_{std}^{2}$
at the minimum is equal to the number of degrees of freedom $\nu$,
here equal to 5 (9 periods minus four free parameters). This leads
to a scale factor $S$ between the standard minimum $\chi_{{\rm std}}^{2}$
and the value of the merit function $\overline{\chi}^{2}$ at the
optimal point in parameter space given by $S=5/0.0784=63.7755$. We
next compute the value of $\Delta\chi_{{\rm std}}^{2}$ that must
be added to $\chi_{{\rm std}}^{2}$ to cover a range of parameter
space sufficient to reach a given confidence level in the estimates
of the uncertainties on the parameters. We adopt the $1\sigma$ limit
(68.3\% confidence level) and compute $\Delta\chi_{{\rm std}}^{2}$
using the \noun{Gammq} routine of \citet{1985nreb.book.....V} for
the truncated gamma function. For the case $\nu=5$, we find that
$\Delta\chi_{{\rm std}}^{2}=5.8907$. Applying the renormalization
according to the scale factor $S$, this corresponds to $\Delta\overline{\chi}^{2}=0.0924$,
the value we now use in equation (\ref{eq:errors}). We finally find
that, at the 68.3\% confidence level, the error estimates associated
with the four primary parameters of \object{PG~1219+534} (i.e., $\log g$,
$T_{{\rm eff}}$, $M_{*}$, and $\log[M_{{\rm env}}/M_{*}]\simeq\log q[{\rm H}]$)
correspond to the values given in Table \ref{cap:Inferred-Properties-of}.
These error estimates are also illustrated as dotted-line rectangles
on the maps provided in Figures \ref{cap:logg-teff} and \ref{cap:lqh-mass}.
Uncertainties on the secondary quantities are then easily derived
from the values obtained for the primary parameters.

As mentioned in the previous paragraph, this approach relies in part
on the assumption that the parameters are independent, an assumption
that is not well verified in practice due, in particular, to the degeneracy
in parameter space of the optimal model solution. To deal with such
a difficulty, one could consider, as an alternative method to provide
estimates of the uncertainties, the relatively strong constraint brought
by spectroscopy on the determination of the effective temperature.
Using projection maps of the merit function such as those illustrated
in Figures \ref{cap:lg-teff-lqh} and \ref{cap:lg-teff-mtot}, one
can delimit the domain of acceptable values for all model parameters
according to the limits (i.e., the evaluated errors) imposed by spectroscopy
on the $T_{{\rm eff}}$ values. A close examination of these maps
indicates, however, that the uncertainties derived from this approach
are of the same order as the values computed from the method introduced
by \citet{2001ApJ...563.1013B}. We therefore maintain our first evaluation
of the errors given in Table \ref{cap:Inferred-Properties-of} as
representative of the $1\sigma$ uncertainties associated with the
derived fundamental parameters of \object{PG~1219+534}.

These asteroseismic results deserve additional comments as they are
representative of the current potential and limitations of asteroseismology
applied to EC14026 pulsators. A first significant contribution of
asteroseismology to the study of sdB stars is the ability to derive
values for the surface gravity $\log g$ with unprecedented accuracy,
i.e., improved by a factor of $\sim10$ compared to current spectroscopic measurements,
typically. This reflects the fact that the $p$-mode periods are particularly
sensitive to the gas density which strongly affects the speed at which
acoustic waves propagate in the stellar interior (i.e, the sound speed).
And how dense and compact a sdB star is largely depends on the $\log g$
parameter (see, e.g., \citealt{2002ApJS..139..487C}). On the other
hand, we find that the asteroseismic value derived for the effective
temperature is not well constrained, due in part to the fact that
the $p$-mode periods are relatively insensitive to this parameter
in sdB stars, but also because solution degeneracies may prevent from
uniquely measuring that quantity through asteroseismology only. Notably,
spectroscopy turns out to be a much more accurate method to measure
$T_{{\rm eff}}$ and allows to lift eventual degeneracies in the asteroseismic
solutions. Therefore, we stress that both methods have indeed complementary
roles to play in this respect. In addition to those {}``classical'' parameters,
asteroseismology allows for the determinations at an interesting level
of accuracy of fundamental structural quantities, such as the total
mass and the thickness of the outer H-rich envelope in hot B subdwarfs,
two key parameters for testing theories of formation and evolution
of stars on the Extreme Horizontal Branch. These two quantities cannot
generally be determined otherwise, except for some masses of sdB stars
known through the study of binary systems containing an sdB component,
but with a relatively poor accuracy on the derived value. For its
part, the determination of the mass of the outer H-rich envelope is
a pure product of asteroseismology and, beside theory, this quantity
cannot be constrained by other known means. Determinations of $M_{{\rm env}}/M_{*}$
for other EC14026 pulsators through our seismic approach may provide,
in the future, interesting insight into the internal structure and
evolution of these old extreme horizontal branch stars.

To conclude this subsection, we point out that we are fully aware
that our asteroseismic determinations of the global parameters of
\object{PG~1219+534} are only as good, in an absolute sense, as the
constitutive physics that went into the construction of the equilibrium
models. Future improvements in our ability to understand and model
sdB stars at the level of the microphysics (i.e., the equation of
state, the opacity, the radiative levitation, initial conditions,
and so on) will necessarily lead to revised estimates of these parameters
and, hopefully, to improved matches of the observed periods. Nevertheless,
we do point out that solid credibility must already be given to current
available constitutive physics since, after successfully reproducing
the 13 observed pulsation periods of \object{PG 0014+067} to better than
$\sim0.8$\% \citep{2001ApJ...563.1013B}, the current models are also
able to explain the presence of 9 pulsation periods in the EC 14026 star
\object{PG~1219+534} with an average accuracy of $\sim0.6$\% on the periods.

\section{Summary and Conclusion}

We observed the luminosity variations of the relatively bright
($V=13.24\pm0.20$), rapidly pulsating hot B subdwarf \object{PG
1219+534} at high sensitivity at the CFHT during a dedicated run in 2004,
March. Such observations were part of an ongoing, long-term project
to monitor the pulsations of EC14026 stars at sufficiently high S/N
ratios to allow for the detection of low-amplitude modes. The objective
is to increase the number of detected modes in known EC14026 stars,
as it constitutes a necessary step for ultimately applying asteroseismic
methods to probe the inner structure of these objects. Our past observing
experience have shown that the ``marriage'' between the Montr\'eal
3-channel photometer \noun{Lapoune} and the CFHT is particularly
efficient for this purpose. For several sdB pulsators observed so
far in the course of this program, we could typically double the number
of mode detections compared to existent data from other sites (see,
\citealt{2001AN....322..387C}).
Our most recent observations along this line made no exception to the
rule and led to the clear identification of nine independent oscillation
periods in \object{PG~1219+534}. This constitutes a significant improvement
over the original data available for this star, which allowed for
the detection of only four dominant periods \citep{1999MNRAS.305...28K}.

From the time series analysis using standard methods that combine
Fourier analysis, least squares fits of the light curve, and prewhitening
methods, we found that \object{PG~1219+534} has a relatively simple
pulsation spectrum characterized by well separated components easily
recognizable as independent modes. Moreover, we found no indication
of close frequency structures (such as multiplets), at least at the
frequency resolution achieved during these observations ($\sim3.44$
$\mu$Hz). This indicates that \object{PG~1219+534} is likely
a slow rotator, with a lower limit on its rotation period that can
be set to $\sim3.4$ days (i.e., the time baseline of our observing
run). The contrast is sharp with two other members of the EC14026 class,
\object{PG 1605+072} \citep{1998MNRAS.296..317K} and
\object{KPD 1930+2752} \citep{2000ApJ...530..441B}, which are rather exceptional in that both show
very complex oscillation spectra due most certainly to rapid rotation.
\object{PG~1219+534} appears instead as a well behaved sdB pulsator
with a rich -- yet simple enough -- period spectrum ideally suited
for detailed asteroseismology. Interestingly, we found that, for this
particular star, one single night of excellent CFHT data turned out to
have all the seismic information necessary for a subsequent successful
asteroseismological exercise. Our detailed analysis of the
light curves of \object{PG~1219+534} also illustrated the fact that,
for EC 14026 pulsators having an easily resolved period spectrum,
the search for a better signal-to-noise ratio is of higher interest
for the sake of asteroseismology than a longer time baseline and/or
coverage. Indeed, we showed that mixing high quality light curves
with low S/N ratio data (due to degraded observing conditions) can
basically ruin all the benefit of having the high sensitivity data,
since the overall noise level in the Fourier domain will be dominated
by the noise introduced by the worst time series included. This has,
of course, implications on how one should treat heterogeneous data
sets such as those gathered during multisite campaigns, for instance.
We mention, along this line, that the best asteroseismic return for
more complex EC14026 pulsators that need multisite contiguous coverage
to unambiguously resolve their pulsation spectrum would require the
acquisition of high sensitivity photometric data which is as homogeneous
as possible in quality. In this context, we are engaged in the
organisations of bi- (or tri-) site campaigns that would involve 4m-class
telescopes positioned at strategic longitudes. Our efforts are presently
directed toward a joint utilisation of the CFHT with \noun{Lapoune}
and the WHT with \noun{Ultracam} as both telescopes are ideally
located geographically and would provide an almost complete coverage.
This, of course, would be in addition to recent efforts to perform
multicolor fast photometry \citep{2004MNRAS.352..699J} and time resolved
spectroscopy \citep{2004Ap&SS.291..457O} on these pulsating stars.

Based on the nine oscillation periods identified in the light curve of
\object{PG~1219+534}, we have attempted a detailed asteroseismic
analysis of this EC14026 star. Our approach relied on the well-known
forward method with the goal of finding objectively the model that
would best match the set of periods observed in this star with a set
of theoretical periods. We have used the second generation
sdB models of \citet{1997ApJ...483L.123C} to compute the theoretical
periods since those have proved to account quite well for the
class properties of the EC14026 pulsators \citep{2001PASP..113..775C}.
Due to the lack of mode identification from the available ``white light''
photometric data, we performed a double-optimization procedure that
takes place simultaneously at the period matching level and in the model
parameter space, leading objectively to the best fit of the asteroseismic
observations. In the present study, we have used the Toulouse package of numerical
tools specifically developed, over the years, for the purpose of asteroseismology
and which includes a genetic algorithm based period matching code (dedicated to the
first optimization step), a parallel genetic algorithm based code, and a parallel
grid computing code allowing for an efficient exploration of the vast parameter
space (the second optimization step).

Our thorough exploration of parameter space led us to first isolate
two families of model solutions that equally best-match the observed
periods of \object{PG~1219+534}. These families of models have surface
gravities around $\log g\sim5.81$ and $\log g\sim5.71$, respectively,
and correspond to slightly different mode identifications. This degeneracy
in the asteroseismic solution could not be lifted solely on the basis
of period fit quality, and additional constraints were required to
select the {}``correct'' one. Such constraints were provided by
the available spectroscopic measurements for the atmospheric parameters
of \object{PG~1219+534}. An ongoing project aimed at improving the
spectroscopic characterization of sdB stars (including both the rapid
and slow pulsators), and which is based on the acquisition of high
resolution, high S/N ratio spectra with the blue spectrograph at the
new MMT, has proved essential in this context. The independently derived
atmospheric parameters for \object{PG~1219+534}, $T_{{\rm eff}}=$
33,600 $\pm$ 370 K and $\log g=5.810\pm0.046$, allowed us to clearly
reject the family of model solutions found at low surface gravity
(i.e., near $\log g\sim5.71$). Concentrating on the preferred solution
located near $\log g\sim5.81$, we found however that another kind
of degeneracy exists in the solution which, again, precludes from uniquely
isolate an optimal model from asteroseismology alone. This problem
occurs due to concurrent effects of model parameters on the pulsation
periods, leading, in the worst case, to a line-degeneracy of the best-fit
solutions that mostly link the total mass parameter with the effective
temperature parameter. We found, indeed, that seismic solutions of
comparable quality exist for all values of the total mass of the star,
but these solutions have values for $T_{{\rm eff}}$ that are correlated.
The trend uncovered is monotonic and indicates that optimal models
of low-mass (down to $0.30$ $M_{\odot}$) require cooler effective
temperatures, while optimal models of high-mass (up to $0.53$ $M_{\odot}$,
but also verified with a few additional calculations at higher masses)
need hotter effective temperatures. It is again spectroscopy that
provided the additional information needed to bypass this difficulty,
as using the well-constrained value of $T_{{\rm eff}}$ derived from
this method led to the determination of a unique viable optimal model
solution that fit consistently \object{PG~1219+534}, both asteroseismologically
and spectroscopically.

The basic properties of \object{PG~1219+534} as inferred from our
combined spectroscopic and asteroseismic approach have been summarized
in Table \ref{cap:Inferred-Properties-of}. The quest for an optimal
seismic model of \object{PG~1219+534} has revealed that accurate
spectroscopy remains an essential ingredient for detailed asteroseismology
of EC 14026 stars in order to converge toward a unique solution. A
relatively precise determination of the surface gravity using spectroscopic
techniques is generally helpful to discriminate between families of
potential seismic solutions. Once the ``right'' family has been
chosen, though, the constraint brought by asteroseismology on the
value of $\log g$ is much tighter than the precision that spectroscopy
can currently provide. This is one of the outstanding outcomes of
asteroseismology as it leads to measurements of the surface gravity of
EC 14026 stars  at unprecedented accuracy (a gain of a factor $\sim10$,
typically). The opposite is true for the effective temperature and the
spectroscopic value for this parameter must clearly be
preferred. Moreover, in the case of \object{PG~1219+534}, the line
degeneracy uncovered for the seismic solution precludes an
independent measurement of $T_{{\rm eff}}$ through asteroseismological
means.

Another outstanding outcome of asteroseismology is the measurement of
the total mass and the mass of the H-rich envelope at an interesting
level of accuracy. In our study of \object{PG~1219+534}, we found
however that the total mass could not be inferred independently of the
spectroscopic measurement of the effective temperature. Nonetheless, the
value derived, $M=0.457\pm0.012$ $M_{\odot}$, indicates a mass of
\object{PG~1219+534} that is close to the canonical value for sdB stars.
Although this value of the mass depends on how
reliable the spectroscopic estimate of $T_{{\rm eff}}$ is (particularly
in regards to eventual systematic effects), we stress that significantly
different masses than the derived -- close to canonical -- value, such
as some scenarios of binary evolution suggest, would require a shift of
several thousand Kelvins in effective temperature to be compatible with
asteroseismology. Such a drastic change in the evaluation of $T_{{\rm
eff}}$ from spectroscopy is highly unlikely and, therefore, we are
confident in concluding that \object{PG~1219+534} has a mass close to
the canonical mass of extreme horizontal branch stars.

We found that our optimal model solution is able to reproduce simultaneously
the 9 periods observed in \object{PG~1219+534} with an average dispersion
of only $\sim0.6$\%. The observed periods correspond to low-order
$\ell=0$, 1, 2, and 3 acoustic modes that are, indeed, predicted
to be excited according to nonadiabatic pulsation theory. It is a
remarkable result (not often achieved so far in the field of
asteroseismology) that the model solution uncovered for \object{PG
1219+534} can, at the same time, reproduce all the periods observed in
this star at a relatively high level of accuracy, be consistent with
nonadiabatic theory, and satisfy the spectroscopic constraints. Such a
consistency between three independent aspects of the modeling of these
pulsating stars is a result that was not guaranteed at the outset. This
is strong indication that the basic constitutive physics which enters in the
construction of our current models used for the asteroseismic analyses
of EC 14026 stars is sound, especially as it follows similar results
already obtained for another rapid sdB pulsator, \object{PG 0014+067} \citep{2001ApJ...563.1013B}.
Furthermore, the excellent agreement between observations and theory
at the nonadiabatic level brings further proof that the iron bump
opacity mechanism of \citet{1997ApJ...483L.123C} is indeed at the
origin of the EC 14026 phenomenon.

Finally, we note that the mode identification resulting from our detailed
asteroseismic analysis of \object{PG~1219+534} constitutes a prediction
that, in principle, can be independently tested using multicolor photometry.
Along this line, we initiated a project to use the Far UV Spectroscopic
Explorer (FUSE) as a fast-photometer (a possibility offered by the
so-called {}``Time-Tag'' mode of the instrument) that would provide
a light curve of \object{PG~1219+534} in the FUV (near $\sim1000$\AA)
bandpass. Combined with nearly simultaneous ground based observations
in the optical, this project should lead to strong independent constraints
on the $\ell$ value of the modes seen in \object{PG~1219+534}, from
which the predicted mode identification reported in this paper will
be checked.

\begin{acknowledgements}
This work was supported in part by the NSERC of Canada and by the Fund
FQRNT (Qu\'ebec). G.F. also acknowledges the contribution of the Canada
Research Chair Program.

\bibliographystyle{aa}
\addcontentsline{toc}{chapter}{\bibname}\bibliography{/home/stephane/Astrophysique/Library/Bibtex/charpinet,/home/stephane/Astrophysique/Library/Bibtex/dorman,/home/stephane/Astrophysique/Library/Bibtex/heber,/home/stephane/Astrophysique/Library/Bibtex/fontaine,/home/stephane/Astrophysique/Library/Bibtex/han,/home/stephane/Astrophysique/Library/Bibtex/kilkenny,/home/stephane/Astrophysique/Library/Bibtex/green-schmidt-liebert,/home/stephane/Astrophysique/Library/Bibtex/saffer,/home/stephane/Astrophysique/Library/Bibtex/wesemael,/home/stephane/Astrophysique/Library/Bibtex/billeres,/home/stephane/Astrophysique/Library/Bibtex/jeffery,/home/stephane/Astrophysique/Library/Bibtex/press,/home/stephane/Astrophysique/Library/Bibtex/bevington,/home/stephane/Astrophysique/Library/Bibtex/schuh}
 \end{acknowledgements}

\end{document}